\pdfoutput=1
\documentclass[11pt, letter,fleqn]{article}
\usepackage{graphics,color}
\usepackage{caption}
\usepackage{subcaption}
\usepackage{cprotect}
\usepackage{epstopdf}
\usepackage{graphicx}
\usepackage{verbatim}
\usepackage{float}
\usepackage{listings}
\usepackage{tikz}
\usepackage[section]{placeins}
\usepackage{multirow}
\usepackage{pdflscape}	
\usepackage{enumitem}  %for nested itemize
\usepackage{cite}

\usepackage{refstyle}
\usepackage{amsmath, amssymb,amsthm}
\usepackage{chngcntr}
\counterwithin{equation}{section}
\newtheorem{prop}{Proposition}

\title{New Exact Plasma Equilibria with Axial and Helical Symmetry}

\author{  Jason M. Keller\footnotemark[1], ~~Alexei
  F. Cheviakov\footnotemark[2] \\ {\small
    \emph{Department of Mathematics and Statistics, University of
      Saskatchewan, Saskatoon, S7N 5E6 Canada}}
}

 {\theoremstyle{definition}

{\theoremstyle{definition} }
{\theoremstyle{definition} }

\def\const{\hbox{\rm const}}

\def\grad{\mathop{\hbox{\rm grad}}}

\def\div{\mathop{\hbox{\rm div}}}
\def\curl{\mathop{\hbox{\rm curl}}}
\def\vec#1{{\boldsymbol{\rm #1}}}  %\def\vec#1{{\bf {#1}}}

\def\beq{\begin{equation}}
\def\eeq{\end{equation}}
\def\barr{\begin{array}{ll}}
\def\earr{\end{array}}

\begin{document}

%\footnotetext[1]{Corresponding author. Electronic mail: jmk810@usask.ca}
%\footnotetext[2]{Work done within NSERC Undergraduate Summer Research %Fellowship.}
%\footnotetext[2]{Electronic mail: cheviakov@math.usask.ca}

\maketitle

\begin{abstract}
Exact solutions to the static equilibrium  magnetohydrodynamic equations are presented and discussed for both axially and helically reduced systems. For both symmetries, physical restrictions on the solutions are discussed and it is seen that there arise two different families of physical solutions corresponding to two distinct pressure profiles. One is suitable for atmospherically confined plasma whereas the other is suitable for plasma residing in a vacuum. For solutions in vacuum, the radial dependence is found to behave periodically for both axial and helical symmetries. For these physical solutions, only a truncated solution inside of some chosen domain can be considered. The magnetic field boundary can than be described by a current sheet. In the last part of the paper, these static equilibrium solutions are transformed into solutions with non-zero fluid velocity.
\end{abstract}

\section{Introduction}
In nature many plasma configurations have different approximate symmetries. The types of symmetries that are most common for plasma include axial symmetry as well as helical symmetry. These symmetries can be exposed in the MHD equations to greatly simplify the pursuit of exact solutions. Some solutions have already been found using these reduction methods and are already discussed in literature \cite{bogoyavlenskij2000astrophysical,bogoyavlenskij2000helically,kaiser1995ball,atanasiu2004analytical}. In most cases, the solutions that have been found and discussed have pressure profiles in which the centre of the plasma is at a lower pressure then the boundary. This is not an adequate model for multiple plasma configurations such as a plasma residing in vacuum. One exact physical solution in which the pressure in the center is higher than the boundary for axial symmetry is constructed and discussed in \cite{atanasiu2004analytical}.
\medskip

The aim for the contributions in this paper are to construct and discuss time independent solutions to the incompressible MHD equations for axial and helical configurations with an emphasis towards physical solutions with pressure profiles higher in the center of the plasma and lower at the boundary. The helical symmetric configurations are not discussed in other literature to the best of the authors knowledge.
\medskip

In part 2, the time dependent, time independent and static equilibrium MHD equations are presented and discussed along with needed requirements for physical solutions. Reduction methods on the static equilibrium MHD equations are shown along with some known exact solutions.
\medskip

Part 3 of the paper is concerned with axially symmetric solutions. The static equilibrium MHD equations are reduced to a well known single partial differential equation, (PDE), known as the Grad-Shafranov equation. Two different families of solutions are constructed and discussed corresponding to two different types of pressure profiles, the first with a higher pressure at the boundary of the plasma with the radial part written in terms of Whittaker functions and the later with a higher pressure in the center of the plasma this time having radial dependence written with the Coulomb Wave functions. For the first family, specific examples are shown including one from \cite{bogoyavlenskij2000astrophysical} as a special case. An example of the second family is also shown, however its radially periodic behaviour requires truncation at some boundary to satisfy the physical requirements.
\medskip

In part 4, the helical reduction on the static equilibrium MHD equations leads to the single PDE known as the Johnson-Frieman-Kulsrud-Oberman (JFKO) equation. Solutions to this equation are found again for two different types of pressure profiles. Both solutions have a radial dependence in which the confluent Heun function arises. Examples of the first type of pressure profile are shown including one from \cite{bogoyavlenskij2000helically} as a special case. A second example is then shown for plasmas confined to vacuum where the radially periodic nature requires truncation at some boundary.
\medskip

In part 5, the transformations by Bogoyavlenskij found in \cite{bogoyavlenskij2002symmetry} are used to transform the static MHD solutions from part 3 and part 4 into time independent MHD solutions.

\section{Ideal plasma equilibrium with axial and helical symmetry}

\subsection{Magnetohydrodynamic systems}

One can consider the isotropic MHD model for an ideal plasma as an extension of the Euler equations with an extra forcing term coming from the electrodynamic Lorentz force. For a quasi-neutral plasma with roughly an equal amount of ions and electrons, this force can be written as $\vec{f} = \vec{J} \times \vec{B}$. This isotropic model of plasma is suitable when the mean free path of plasma particles is much less than the typical length scale of the problem. The plasma is also considered ideal with infinite conductivity and negligible viscosity. Both of these approximations are valid for large magnetic and kinetic Reynolds numbers \cite{cheviakov2004symmetries}. Following these assumptions, the ideal isotropic incompressible MHD model takes on the form

\begin{subequations}\label{eq:MHD}
\begin{equation}
\frac{\partial \rho}{\partial t} + \div \,\rho\vec{V} = 0,
\end{equation}
\begin{equation}
\rho \frac{\partial \vec{V}}{\partial t} = \rho\vec{V}\times \curl \vec{V} + \vec{J} \times \vec{B} -\grad P - \rho \grad\frac{\vec{V}^2}{2},
\end{equation}
\begin{equation}
\frac{\partial \vec{B}}{\partial t} = \curl(\vec{V} \times \vec{B}),
\end{equation}
\begin{equation}
\div \vec{B} = 0, \quad \div \vec{V} = 0,
\end{equation}
\end{subequations}

where $\vec{J} = \frac{1}{\mu} \curl \vec{B}$ is the electric current density. In (\ref{eq:MHD}) $\rho$ is the plasma density, $\vec{V}$ is the plasma velocity, $P$ is the scalar pressure (models of anisotropic plasma's utilize tensor valued pressure), $\vec{B}$ is the magnetic induction vector and $\mu$ is the magnetic permeability of the plasma.\medskip

First considering the equilibrium condition for (\ref{eq:MHD}), when this system is time independent ($\partial/\partial t = 0$), (\ref{eq:MHD}) then takes on the form
\begin{subequations}\label{eq:MHDeq}
\begin{equation}\label{eq:MHDeqCont}
\div {\rho \vec{V}} = 0,
\end{equation}
\begin{equation}
\rho \vec{V} \times \curl \vec{V} + \vec{J} \times \vec{B} - \grad{P} - \rho \grad{\frac{\vec{V}^2}{2}} = 0,
\end{equation}
\begin{equation}\label{eq:MHDtop}
\curl{(\vec{V} \times \vec{B})} = 0,
\end{equation}
\begin{equation}
\div \vec{B} = 0, \quad
\end{equation}
\begin{equation}\label{eq:MHDeqState}
\div \vec{V} = 0.
\end{equation}
\end{subequations}

An immediate consequence of \eqref{MHDeqCont}, \eqref{MHDeqState} is
\begin{equation}{\label{eq:FIELD_align}}
(\grad \rho) \cdot \vec{V} = 0,
\end{equation}
which implies that the plasma density does not change along streamlines.\medskip

A further reduction of (\ref{eq:MHD}), (\ref{eq:MHDeq}) is the static equilibrium MHD system ($\vec{V} = 0$) given by
\begin{subequations}\label{eq:MHDst}
\begin{equation}
\div \vec{B} = 0,
\end{equation}
\begin{equation}
\curl \vec{B} \times \vec{B} = \mu \grad P.
\end{equation}
\end{subequations}
Using the typical scales of space, magnetic field strength, and pressure,
\[
x=Lx^*,\quad y=Ly^*,\quad z=Lz^*,\quad P=P_0 P^*,\quad B_i=B_0B_i^*,\quad i=1,2,3,
\]
with $P_0, B_0, L=\const$ with $P_0=B_0^2/\mu$, and omitting asterisks, the system of PDEs \eqref{MHDst} is converted to the same equations with $\mu=1$, which will be assumed below. \\\\
The independent and dependent variables in (\ref{eq:MHDst}) are respectively considered dimensionless.
For the purposes of this paper, several coordinates systems will be utilized; in Cartesian coordinates $(x,y,z)$, cylindrical coordinates $(r,\varphi,z)$, and helical coordinates $(r, \eta, \xi)$, the magnetic field is then given by
\begin{equation}
\vec{B}= B_1\vec{e}_x+B_2\vec{e}_y+B_3\vec{e}_z = B_r\vec{e}_r+B_\varphi\vec{e}_\varphi+B_3\vec{e}_z.
\end{equation}

and the helical coordinates can be defined from cylindrical coordinates as
\begin{equation}\label{eq:helical_coordinates}
r = r, \quad \eta = \varphi - \gamma z/r^2, \quad \xi = z - \gamma\varphi,
\end{equation}
where $\gamma$ is some constant parameter.
In addition to the regularity and sufficient smoothness of the dependent variables $\vec{B}, \vec{V}, P, \rho$ in \eqref{MHDst}, other natural requirements for a physically relevant solution of (\ref{eq:MHDeq}) in an isolated plasma domain $\mathcal{V}$ not sustained by external forces include
\begin{enumerate}\label{Physical Constraints}
  \item the vanishing pressure condition: $P=0$ outside $\mathcal{V}$, or $P\leq P_0=\const>0$ with $P\to P_0$ when $|\vec{x}|\to\infty$,
  \item finite magnetic energy: $\displaystyle{\int_{\mathcal{U}} |\vec{B}(\vec{x})|^2\, d^3 x <+\infty}$.
\end{enumerate}
Here $\mathcal{U}$ is either the whole plasma domain, $\mathcal{U}=\mathcal{V}$, or, for a plasma configuration stretched along, say, the $z$-axis, a slice $x,y\in \mathbb{R}$, $z_1\leq z\leq z_2$.
\medskip

For a plasma in a domain $\mathcal{U}$ bounded by $\partial \mathcal{U}$, in which $\vec{B}, \vec{V}, P \to 0$ on $\partial \mathcal{U}$, the tangential boundary of the magnetic field can be described with the following boundary condition
\begin{equation}\label{eq:Boundary_Condition}
\vec{H} \times \vec{n} = \vec{K} \times \vec{n},
\end{equation}
where
\begin{equation*}
\vec{H} = \frac{\vec{B}}{\mu}.
\end{equation*}
Here $\partial \mathcal{U}$ is referred to as the \emph{current sheet}, $\vec{n}$ is the outward facing normal and $\vec{K}$ is referred to as the \emph{surface current density}.
\medskip

An important property of equilibrium (\ref{eq:MHDeq}) and static equilibrium (\ref{eq:MHDst}) MHD models are their field line topology. Some important applications of these properties can be found in \cite{bogoyavlenskij2002symmetry, cheviakov2004symmetries}. Specifically, in most cases, the plasma streamlines and magnetic field lines are tangent to \emph{magnetic surfaces} spanning the plasma domain. For MHD equilibrium (\ref{eq:MHDeq}), if $\vec{V}$ and $\vec{B}$ are non-parallel, it follows from \eqref{MHDtop} that locally there exists a function $\alpha = \alpha(\vec{x})$ such that $\vec{V} \times \vec{B} = \grad \alpha$, and thus, $\alpha = \const$ defines the magnetic surfaces. In the case when $\vec{V}$ and $\vec{B}$ are parallel, (i.e. $\vec{V} = f(\vec{x})\vec{B}$), which, from the solenoidality of $\vec{V}$ and $\vec{B}$ one has $\grad f(\vec{x})\cdot \vec{V} = 0$, so both $f(\vec{x})$ and $\rho(\vec{x})$, from \eqref{FIELD_align}, are constant on plasma streamlines and magnetic field lines. Similarly, for static equilibrium (\ref{eq:MHDst}), if $\curl \vec{B}$ and $\vec{B}$ are non-parallel, then the pressure $P$ will be non-constant, and for lines of constant pressure the magnetic surfaces will be enumerated. Lastly, for $\curl \vec{B}$ and $\vec{B}$ parallel (i.e. $\curl\vec{B} = \alpha(\vec{x})\vec{B}$) both $\curl \vec{B}$ and $\vec{B}$ are tangent to magnetic surfaces which are enumerated by constant levels of $\alpha(\vec{x})$. One can conclude that all magnetic field lines and plasma streamlines, except for Beltrami flows ($\alpha(\vec{x}) = \const$), lie on 2D magnetic surfaces and for uniquely defined magnetic surfaces these surfaces will be topologically equivalent to tori \cite{cheviakov2004symmetries}.

\subsection{Symmetry reductions of static equilibrium MHD equations}

In nature several types of stable plasma equilibrium configurations are observed to approximately exhibit axial and helical symmetries. In particular, such approximations have been used in modelling of plasma in tokamak fusion reactors \cite{PhysRevLett.93.155007,4694261}. In this confinement, the plasma is axially symmetric. These observed symmetries act as the motivation to seek symmetric solutions of static equilibrium equations.
\medskip

In the case of axial symmetry, one has $\vec{B} = \vec{B}(r,z)$, $P = P(r,z)$ in which \eqref{MHDst} can be reduced to a single PDE after using the axisymmetric magnetic field $B(r,\phi,z) = \nabla \phi \times \psi(r,z) + I(\psi)\nabla \phi$ written in terms of the flux function $\psi(r,z)$ which is independent of $\phi$. This PDE is known as the Grad-Shafranov equation \cite{kaiser1995ball}.

\begin{equation}\label{eq:GS}
\frac{\partial^2 \psi}{\partial r^2} + \frac{\partial^2 \psi}{\partial z^2} - \frac{1}{r}\frac{\partial \psi}{\partial r} + I(\psi)I'(\psi) = -r^2 P'(\psi).
\end{equation}
Here, the magnetic field $\vec{B}$ is expressed in terms of $\psi$ as
\begin{equation}\label{eq:Axial_Magnetic_Field}
\vec{B} = \frac{\psi_z}{r}\vec{e}_r + \frac{I(\psi)}{r}\vec{e}_\phi - \frac{\psi_r}{r}\vec{e}_z,
\end{equation}
$P(\psi)$ is the plasma pressure that is functionally dependent on the unknown flux function, and $I(\psi)$ is an arbitrary function describing the toroidal magnetic field. In \eqref{GS}, the primes denote usual derivatives.
\medskip

Various particular exact solution of the Grad-Shafranov \eqref{GS} have been derived over the years, including solutions in spherical coordinates \cite{kaiser1995ball,bobnev} modelling ball-like plasma configurations. Other interesting solutions to the Grad-Shafranov \eqref{GS} are found in \cite{bogoyavlenskij2000astrophysical,atanasiu2004analytical}. In \cite{bogoyavlenskij2000astrophysical}, Bogoyavlenskij was able to write the following solution to \eqref{GS} corresponding to axially symmetric plasma jets after choosing $P = p_0 - 2\beta^2\psi^2$ and $I = \alpha \psi$.
\begin{equation}\label{eq:axial_jets}
\Psi(r,z) = e^{-\beta r^2}\left(a_N L^*_N(2 \beta r^2) + \sum_{n=1}^{N-1}a_n\sin(\omega_n z + b_n)L^*_n(2\beta r^2)\right).
\end{equation}
Here $L_n^*$ are primitive functions of the Laguerre polynomials and $\omega_n = \sqrt{8\beta(N-n)}$.
\subsubsection{Helically symmetric static equilibrium reduction}
Similarly, after imposing helical symmetry on \eqref{MHDst}, this system of PDEs can be reduced to a single PDE in the two helical coordinates $(r, \xi)$ as derived in \cite{johnson1958some} and refereed to as the JFKO equation
\begin{equation}\label{eq:JFKO}	
\frac{\Psi_{\xi \xi}}{r^2} + \frac{1}{r} \frac{\partial}{\partial r} \bigg{(} \frac{r}{r^2 + \gamma^2} \Psi_r \bigg{)} + \frac{ I(\Psi) I'(\Psi)}{r^2+ 	\gamma^2} + \frac{2 \gamma I(\Psi)}{(r^2+ \gamma^2)^2} = -\mu P'(\Psi).
\end{equation}
Here subscripts denote partial derivatives and primes denote regular derivatives with respect to the given variable. Again, $P(\psi)$ and $I(\psi)$ are arbitrary functions of $\psi$. The magnetic field $\vec{B}$ written in cylindrical coordinates has the following form
\begin{equation}
\vec{B} = \frac{\psi_\xi}{r}\vec{e}_r +\frac{rI(\psi) + \gamma\psi_r}{r^2 + \gamma^2}\vec{e}_\phi   + \frac{\gamma I(\psi) - r\psi_r}{r^2 + \gamma^2}\vec{e}_z.
\end{equation}
Some exact physical solutions to the helical reduction of \eqref{MHDst} can be found in \cite{helicalexample2} as well as exact solutions to \eqref{JFKO} presented in \cite{bogoyavlenskij2000helically}. These solutions describe helically symmetric astrophysical jets with the exact solutions written as
\begin{equation}\label{eq:helical_jets}
\Psi_{Nmn}(r,\xi) = e^{-\beta r^2}\left(a_N B_{0N}(2\beta r^2) + r^m B_{mn}(2\beta r^2)(a_{mn}\cos(m {\xi}/{\gamma}) + b_{mn}\sin(m \xi/\gamma))\right),
\end{equation}
Here $B_{mn}$ are related to the Laguerre polynomials by the following.
\begin{equation}
B_{mn}(x) = \frac{d^m}{dx^m}L_{m+n} - k_{mn}(x)\frac{d^{m+1}}{dx^{m+1}}L_{m+n}(x)
\end{equation}
Where $k_{mn} = (m+2\beta \gamma^2 - \alpha \gamma)/(4n\beta\gamma^2)$. In \eqref{helical_jets} $N$,$m$,$n$ are arbitrary non-negative integers which all satisfy the inequality $2N>2n+m$ \quad \cite{bogoyavlenskij2000helically}.

\section{Exact axially symmetric plasma equilibria}\label{axial}
There are certain choices of the arbitrary functions $P$ and $I$  in which the Grad-Shafranov \eqref{GS} becomes linear. Choosing the pressure to be quadratic and the arbitrary function related to the poloidal current to be linear:
\begin{equation}\label{eq:P_I_axial_linear}
P(\psi) = P_0 + b\psi + \frac{1}{2}a\psi^2, \quad
I(\psi) = \alpha \psi.
\end{equation}
The Grad-Shafranov equation becomes a linear second order PDE:
\begin{equation}
\frac{\partial^2 \psi}{\partial r^2}  - \frac{1}{r}\frac{\partial \psi}{\partial r} + \frac{\partial^2 \psi}{\partial z^2} +(\alpha^2+ar^2) \psi = -b.
\end{equation}
The homogeneous case, $b = 0$ will be considered,
\begin{equation}\label{eq:GS_lin}
\frac{\partial^2 \psi}{\partial r^2}  - \frac{1}{r}\frac{\partial \psi}{\partial r} + \frac{\partial^2 \psi}{\partial z^2} +(\alpha^2+ar^2) \psi = 0.
\end{equation}
Several interesting solutions to \eqref{GS_lin} have been found through the method of separation of variables. One of these examples was discussed above regarding axial symmetric astrophysical jets with the exact solution given by \eqref{axial_jets}.
The linear homogeneous PDE \eqref{GS_lin} admits separated solutions $\psi(r,z)= R(r)Z(z)$, satisfying
\begin{equation}\label{eq:lin_GS_Separated}
Z''=\lambda Z, \quad  R''-\dfrac{1}{r}R' + (\alpha^2 + ar^2+\lambda)R=0,
\end{equation}
where $\lambda$ is an arbitrary separation constant.
\medskip

Depending on the value of $a$ in the pressure term, one obtains two different families of solutions. The two families of solutions correspond to two different types of pressure profiles. For $a<0$, $P\leq P_0=\const>0$ with $P\to P_0$ when $|\vec{x}|\to\infty$. For the case when $a>0$, $P>0$ inside of $\mathcal{V}$ and outside $P = 0$.

\subsection{The first family of axially symmetric solutions}
The first family of solutions arises when $a=-q^2<0$. This corresponds to a pressure profile, $P(\psi) = P_0 - \frac{1}{2}q^2\psi^2$, which is bounded above by some $P_0$ where $P \to P_0$ when $|\vec{x}| \to \infty$. This model is more appropriate with plasmas confined to atmosphere or in some ambient medium.
\medskip

The case when the separation constant is a negative value $\lambda = -k^2$, $k > 0$, corresponds to a plasma jet model extended along the $z$ - axis \cite{bogoyavlenskij2000astrophysical}. From \eqref{lin_GS_Separated} one has
\begin{equation}\label{eq:Z_solution1}
Z = C_1\sin(kz) + C_2\sin(kz).
\end{equation}
In order to solve the radial differential equation from \eqref{lin_GS_Separated}, the substitution $x = qr^2$ can be used to transform this equation into the more familiar form
\begin{equation}\label{eq:R_problem1}
R'' + \left(-\frac{1}{4} + \frac{\alpha^2 - k^2}{4qx}\right)R = 0.
\end{equation}
This equation is now related to the Whittaker ODE
\begin{equation}\label{eq:Whittaker_ODE}
y''(s)+\left(-\dfrac{1}{4}+\dfrac{\delta}{s}+\dfrac{{1}/{4}-\nu^2}{s^2}\right) y(s)=0,
\end{equation}
with the general solution
\[
y(s) = C_1W_M(\delta, \nu, s) + C_2W_W(\delta, \nu, s).
\]
This gives the following general solution to \eqref{R_problem1}
\begin{equation}\label{eq:Whittaker_Solution}
R(r) = C_1W_M\left(\delta, \frac{1}{2},  qr^2\right)+C_2W_W\left(\delta, \frac{1}{2},  q r^2\right),
\end{equation}
where
\begin{equation}\label{eq:delta}
\delta =\frac{\alpha^2 - k^2}{4q}.
\end{equation}
The separated solution of $\psi$ is thus
\begin{equation}\label{eq:Sep_solution1}
\psi_k(r,z) =  \left(C_1W_M\left(\delta, \frac{1}{2},  q r^2\right)+C_2W_W\left(\delta, \frac{1}{2},  q r^2\right)\right)(C_3\sin kz + C_4\cos kz).
\end{equation}
Where $C_1$, $C_2$, $C_3$ and $C_4$ are all arbitrary constants. For $\delta \geq 1$, (\ref{eq:Sep_solution1}) produces physical solutions. It should be noted that due to the linearity of \eqref{GS_lin}, any linear combination of functions (\ref{eq:Sep_solution1}) yield solutions to \eqref{GS_lin}, including

\begin{equation}\label{eq:General_Solution1}
\Psi(r,z) = \int_{-\infty}^{\infty}\psi_k(r,z)\;dk
\end{equation}
where $C_i = C_i(k)$, $i = 1,2,3,4$ are arbitrary distributions. It should be noted that for $C_i$ being
\begin{equation}
C_1(k) =  \sum_{n=1}^{N-1} \tilde{a_n} \Delta \left(k - \sqrt{\alpha^2 - 4qn}\right), \quad C_2(k) = 0, \quad C_3 = a_n\cos b_n, \quad C_4 = a_n \sin b_n,
\end{equation}
where $\Delta(\vec{x})$ denotes the Dirac delta function, \eqref{General_Solution1} will have a similar form to Bogoyavlenskij's solutions given by \eqref{axial_jets}. This corresponds with \eqref{Sep_solution1} having $\delta \in \mathbb{N}$. The following proposition holds.
\medskip

\begin{prop}
The \eqref{Sep_solution1} is a physical solution without truncation, if and only if $\delta \in \mathbb{N}$.
\end{prop}
The proof of this proposition is given in appendix \ref{appendix:a}.

\subsubsection{Examples of the first family of axially symmetric solutions}
\begin{figure}[h]
	\centering
	\begin{subfigure}{0.32\textwidth}
		\centering
		\includegraphics[width=\textwidth]{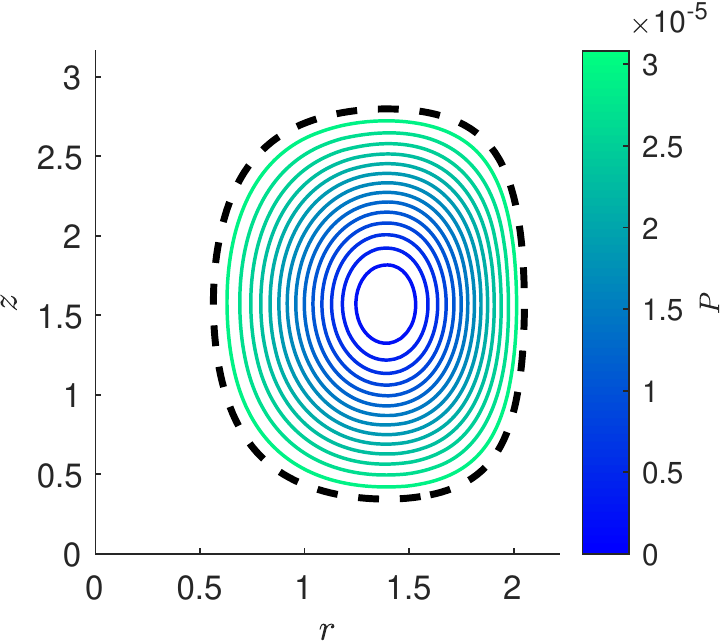}
		\caption{}\label{fig:axial_whittaker_contour_pressure}
		\label{fig:y equals x}
	\end{subfigure}
	\hfill
	\begin{subfigure}{0.32\textwidth}
		\centering
		\includegraphics[width=\textwidth]{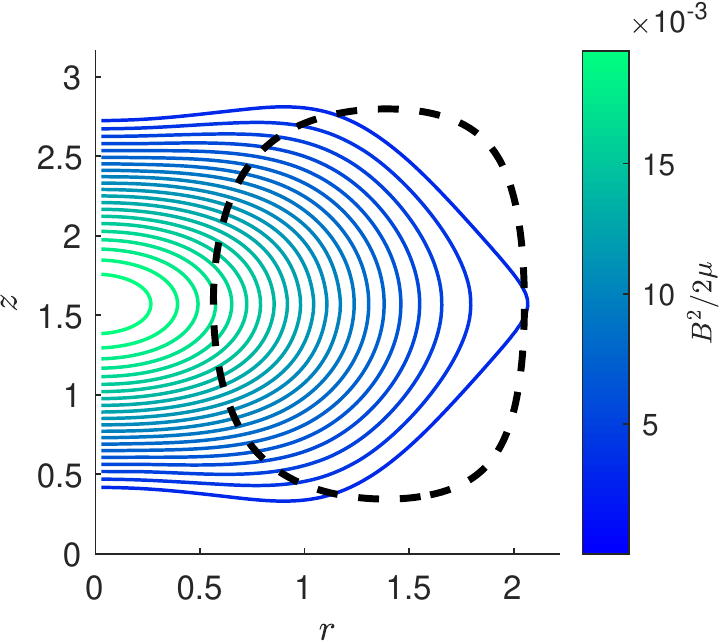}
		\caption{}\label{fig:axial_whittaker_contour_mag_energy}
	\end{subfigure}
	\hfill
	\begin{subfigure}{0.32\textwidth}
		\centering
		\includegraphics[width=\textwidth]{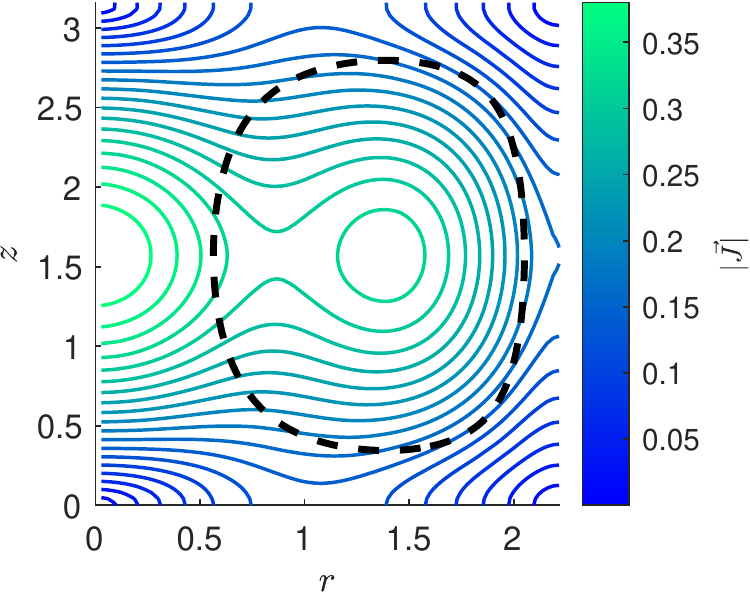}
		\caption{}
		\label{fig:Axial_Whittaker_contour_current_density}
	\end{subfigure}
	\caption{In Figure \ref{fig:axial_whittaker_contour_pressure}, a cross-section of magnetic surfaces $P = \const$ for an axially symmetric plasma equilibrium solution belonging to Family 1 with $\psi$ given by (\ref{eq:Sep_solution1}), for $C_1 = 1$, $C_2 = 0$, $C_3 = 1$, $C_4 = 0$, $\mu = 1$, $q = 0.1$, $\alpha = 2$, $k = 1$ and truncated at the surface $P_0 = 3.3\times 10^{-5}$ (shown with the black dashed line) can be seen. The color-bar shows the values of the dimensionless pressure $P = P_0 -  q^2 \psi^2/2$.  chosen such that $P > 0$ inside of the chosen domain. The corresponding magnetic energy density, $|\vec{B}|^2/2\mu$, can be seen in \ref{fig:axial_whittaker_contour_mag_energy} along with the magnitude of the current density in Figure \ref{fig:Axial_Whittaker_contour_current_density}.}\label{fig:axial_fam1}
\end{figure}

Two different examples are considered for the first solution family. The first utilizes the separated solution \eqref{Sep_solution1} for a case when $\delta$ is not a positive integer. Figure \ref{fig:axial_whittaker_contour_pressure}. It should be noted that in order to achieve finite magnetic energy the domain of this solution needs to be restricted, with the physical boundary involving a current sheet (\ref{eq:Boundary_Condition}). The magnetic energy density is shown in \Figref{axial_whittaker_contour_mag_energy} where the components of the magnetic field are given by \eqref{Axial_Magnetic_Field}. It should be noted that the contours of  \Figref{axial_whittaker_contour_pressure} and \Figref{axial_whittaker_contour_mag_energy} do not coincide. The dotted bold line in \Figref{axial_whittaker_contour_mag_energy} mark the boundary of the plasma domain, and the scalar magnetic energy density plot is only relevant inside of this domain.
\medskip

\begin{figure}[h]
	\centering
	\begin{subfigure}{0.48\textwidth}
		\centering
		\includegraphics[width=\textwidth]{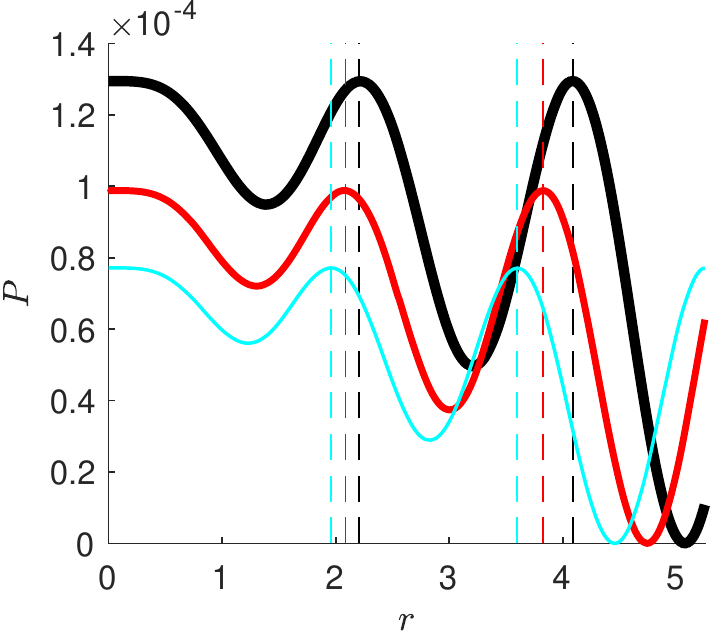}
		\caption{}\label{fig:axial_whittaker_pressure_plot}
	\end{subfigure}
	\hfill
	\begin{subfigure}{0.48\textwidth}
		\hfill
		\centering
		\includegraphics[width=\textwidth]{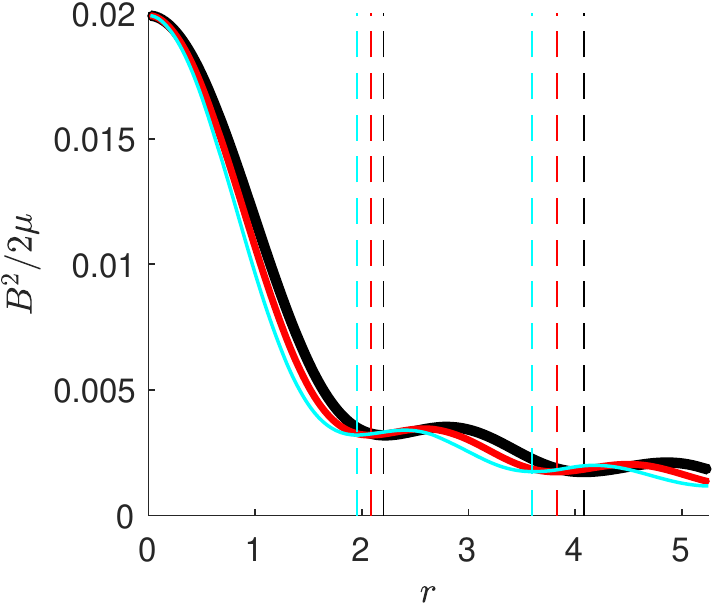}
		\caption{}
		\label{fig:Axial_Whittaker_mag_energy_plot}
	\end{subfigure}
	\caption{A pressure plot of the solution from Figure \ref{fig:axial_fam1} for $z = 1.5$ and $\alpha_i = 2 + 0.1i, i = 1,2,3$ corresponding to the black, red and cyan plots can be seen in Figure \ref{fig:axial_whittaker_pressure_plot}. The different magnetic energy densities for $z = 1.5$ can be seen in Figure \ref{fig:Axial_Whittaker_mag_energy_plot}. The vertical dashed lines for each colour correspond to truncation surfaces that give physical solutions.}\label{fig:Axial_fam1_plot}
\end{figure}

 The second example that will be shown for this first family of axial solutions is linear combination of the special case of \eqref{Sep_solution1} when $\delta$ given by (\ref{eq:delta}) takes on a positive integer value producing Whittaker functions related to the Laguerre polynomials. In this case, it can be considered a global solution as both $P, B^2/2 \to 0$ as $r \to \infty$ satisfying both of the physical constraints. This is the same type of solution as those discussed in \cite{bogoyavlenskij2000astrophysical} with the solution's formula written by \eqref{axial_jets}. A cross-section of the magnetic surfaces can be see in \Figref{axial_whittaker_contour_pressure} along with a contour of the magnetic energy density seen in \Figref{axial_whittaker_contour_mag_energy}. It can be seen with this example that the magnetic energy is concentrated about the center of the plasma.
\medskip

\begin{figure}[htbp]
\begin{center}
\includegraphics[width=.7\textwidth]{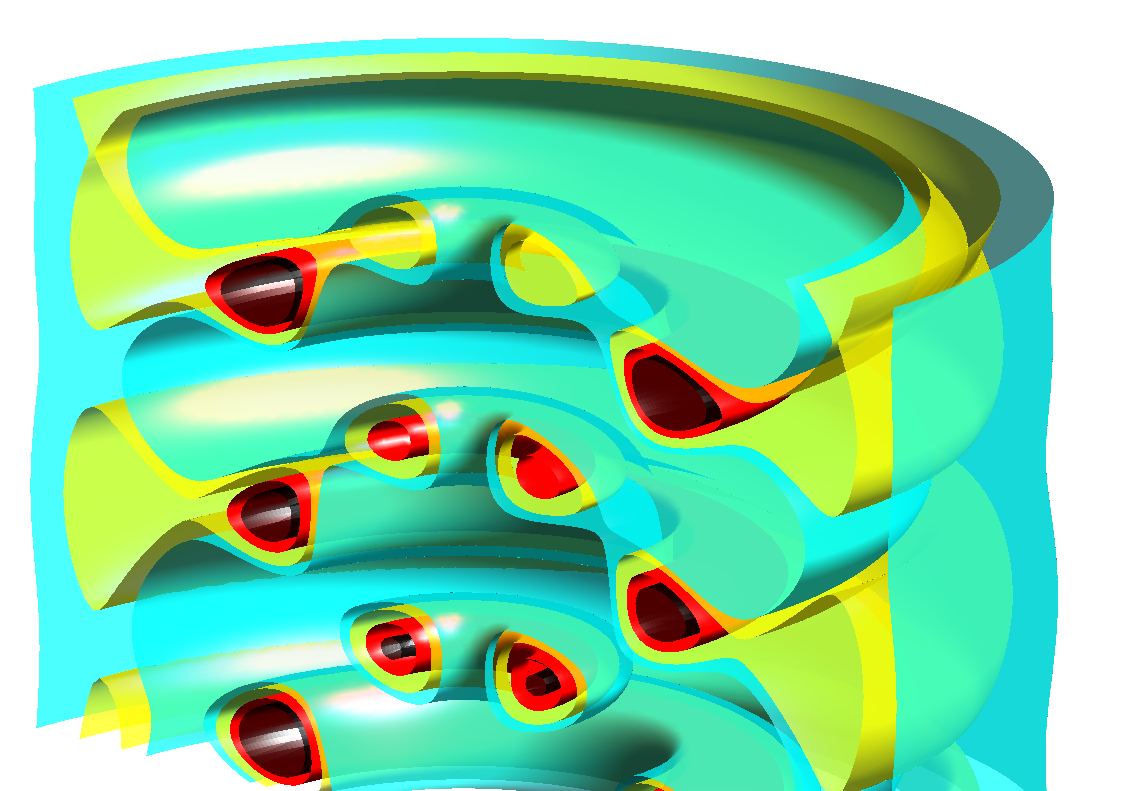}
\end{center}
\caption{\label{fig:Axial_Bog_surf}Axially symmetric magnetic surfaces $\psi, P=\const$ for special case of Family 1, \eqref{axial_jets}, for N = 3, $\beta = 0.1$, $\alpha^2 = 24\beta$. Takes on the form of nested tori and wavy cylinders.}
\end{figure}

With the rotation of \Figref{axial_whittaker_contour_pressure} about the $z$ axis, the 3D magnetic surfaces to which both $\vec{B}$ and $\curl \vec{B}$ are tangent can be seen. This surface is shown in \Figref{Axial_Bog_surf}.

\begin{figure}[h]
	\centering
	\begin{subfigure}{0.32\textwidth}
		\centering
		\includegraphics[width=\textwidth]{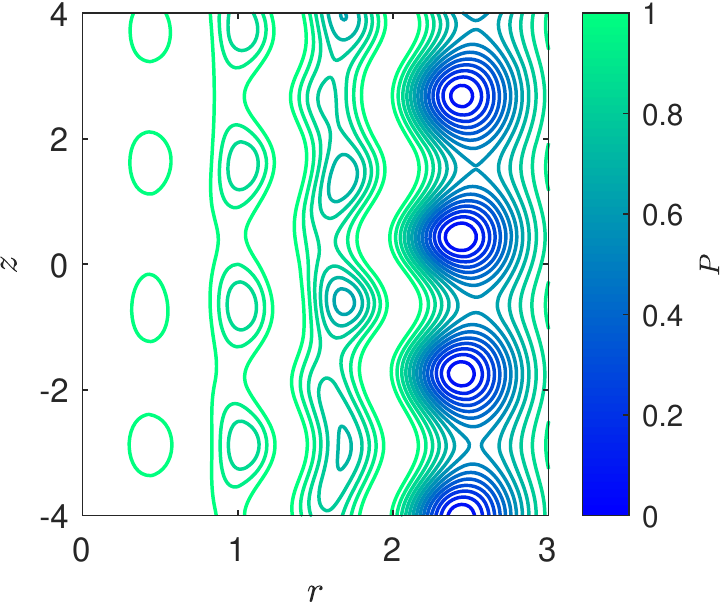}
		\caption{}\label{fig:whittaker_bog_contour_pressure}
		\label{fig:y equals x}
	\end{subfigure}
	\hfill
	\begin{subfigure}{0.32\textwidth}
		\centering
		\includegraphics[width=\textwidth]{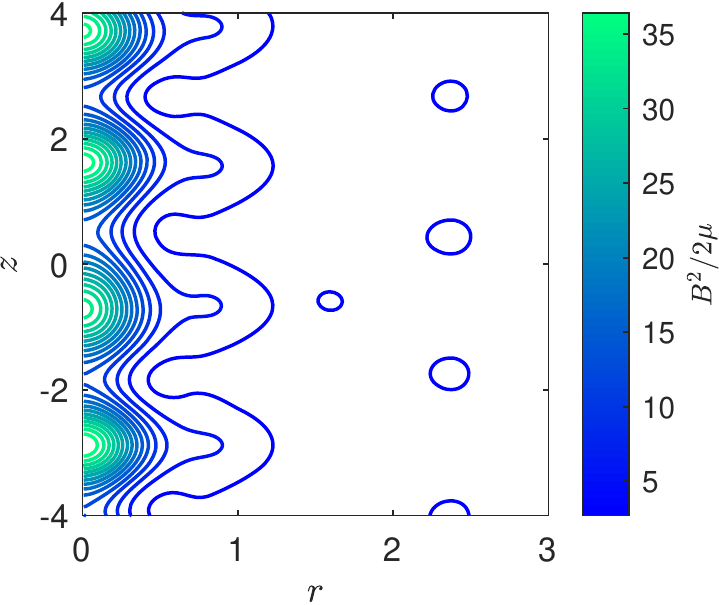}
		\caption{}\label{fig:whittaker_bog_contour_mag_energy}
	\end{subfigure}
	\hfill
	\begin{subfigure}{0.32\textwidth}
		\centering
		\includegraphics[width=\textwidth]{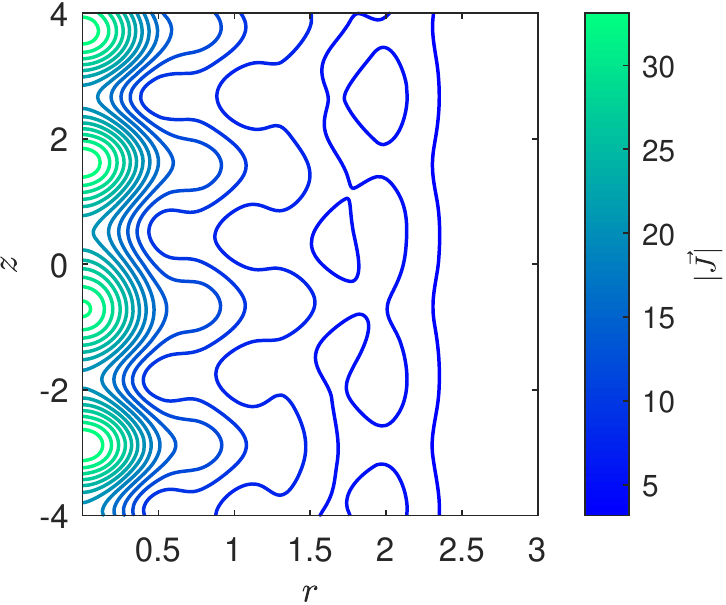}
		\caption{}
		\label{fig:Whittaker_bog_contour_current_density}
	\end{subfigure}
	\caption{In Figure \ref{fig:whittaker_bog_contour_pressure}, a cross-section of magnetic surfaces $P = \const$ for an axially symmetric plasma equilibrium solution belonging to Family 1 for the case when $\delta \in \mathbb{N}$. The color-bar shows the values of the dimensionless pressure $P = P_0 -  q^2 \psi^2/2\mu$ in Figure \ref{fig:whittaker_bog_contour_pressure}. The corresponding magnetic energy density, $|\vec{B}|^2/2\mu$, can be seen in \ref{fig:whittaker_bog_contour_mag_energy} along with the magnitude of the current density in Figure \ref{fig:Whittaker_bog_contour_current_density}.}\label{fig:axial_fam1_deltaint}
\end{figure}

\begin{figure}[h]
	\centering
	\begin{subfigure}{0.48\textwidth}
		\centering
		\includegraphics[width=\textwidth]{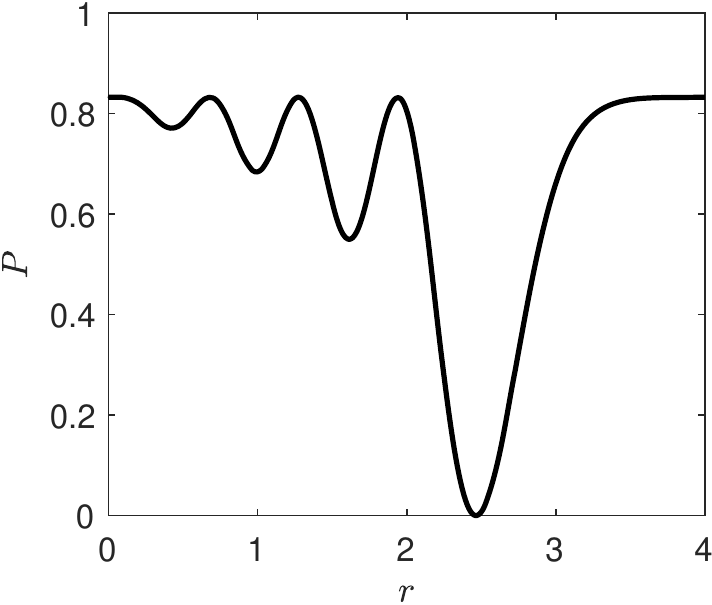}
		\caption{}\label{fig:whittaker_bog_pressure_plot}
	\end{subfigure}
	\hfill
	\begin{subfigure}{0.48\textwidth}
		\hfill
		\centering
		\includegraphics[width=\textwidth]{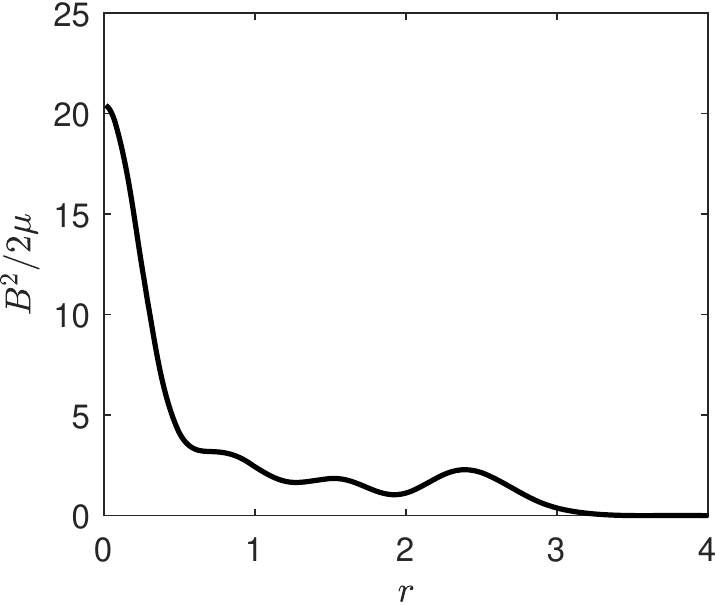}
		\caption{}
		\label{fig:whittaker_bog_mag_energy_plot}
	\end{subfigure}
	\caption{A pressure plot of the solution from Figure \ref{fig:axial_fam1} for $z = 0$ can be seen in \ref{fig:whittaker_bog_pressure_plot}. The magnetic energy densities can be seen in Figure \ref{fig:Axial_Whittaker_mag_energy_plot}. Here we can see that the majority of the magnetic energy is focused around the center of the plasma.}\label{fig:Axial_fam1_plot}
\end{figure}

\subsection{The second family of axially symmetric solutions}
This second family of solutions arises when $a = q^2 >0$. This corresponds to plasmas residing in vacuum in which the pressure, $P = 0$ outside of $\mathcal{V}$ with $P > 0$ inside the domain. Starting from the separated equations given by (\ref{eq:lin_GS_Separated}) with $a = q^2$, the same plasma jet ansatz for $Z(z)$ given by \eqref{Z_solution1} can be used. After transforming the radial problem from (\ref{eq:lin_GS_Separated}) with $x = iqr^2$, where $i$ is the imaginary unit, the following ODE related to the Whittaker ODE \eqref{Whittaker_ODE} is obtained.

\begin{equation}\label{eq:R_problem2}
R'' + \left(-\frac{1}{4} + \frac{k^2 - \alpha^2}{4qx}i\right)R = 0.
\end{equation}

Therefore, by utilizing the general solution to the Whittaker ODE \eqref{Whittaker_ODE}, the radial solution can be written in terms of Whittaker functions of a complex argument and parameter:

\begin{equation}
R(r) = C_1W_M\left(-i\delta, \frac{1}{2}, iqr^2\right)+C_2W_W\left(-i\delta, \frac{1}{2}, iqr^2\right),
\end{equation}
where $\delta$ is given by \eqref{delta}.
\medskip

A relationship exists between Whittaker functions and the Coulomb wave functions:
\begin{equation}\label{Whit_to_Coulomb}
W_M\left(-i\delta, \frac{1}{2}, iqr^2\right) =\frac{2\,i\,\mathcal{C}_F(0,-\delta, qr^2/2)}{|\Gamma(1-i\delta)|e^{\frac{\pi}{2}-\delta}}.
\end{equation}
This gives the motivation to transform the radial problem given in (\ref{eq:lin_GS_Separated}) into a related version of the Coulomb wave ODE using $x = qr^2/2$,
\begin{equation}\label{eq:R_coulomb}
R'' + \left(1 + 2\frac{\alpha^2 - k^2}{4qx}\right)R = 0.
\end{equation}
which is related to the Coulomb wave ODE
\begin{equation}\label{eq:Coulomb_ODE}
y''(s) + \left(1 - \frac{2\sigma}{s} - \frac{L(L + 1)}{s^2}\right)y(s) = 0,
\end{equation}
with the general solution $y(s) = C_1\,\mathcal{C}_F(L,\sigma,s) + C_2\,\mathcal{C}_G(L,\sigma,s)$. By comparing (\ref{eq:R_coulomb}) with (\ref{eq:Coulomb_ODE}) and substituting back $x = qr^2/2$, one arrives at
\begin{equation}
R(r) = C_1\,\mathcal{C}_F\left(0,-\delta,\frac{q}{2}r^2\right) + C_2\,\mathcal{C}_G\left(0, -\delta,\frac{q}{2}r^2\right).
\end{equation}
The second family of solutions to \eqref{GS_lin} corresponding to plasma confined in a vacuum  is:
\begin{equation}\label{eq:Sep_solution_2}
\psi_k(r,z) = \left(C_1\,\mathcal{C}_F\left(0, -\delta,\frac{q}{2}r^2\right) + C_2\,\mathcal{C}_G\left(0, -\delta,\frac{q}{2}r^2\right)\right)(C_3\sin kz + C_4\cos kz),
\end{equation}
where $\delta$ is the same as the first family, given by \eqref{delta}.
Again, any linear combination of the above separated solution is also a solution which can be written as
\begin{equation}\label{eq: General Solution2}
  \Psi(r,z) = \int_{-\infty}^{\infty}\psi_k(r,z)\;dk,
\end{equation}
assuming $C_i = C_i(k)$, $i = 1,2,3,4$. It should be noted that \eqref{Sep_solution_2} does not globally satisfy all of the physical constraints, as both $\mathcal{C}_F$ and $\mathcal{C}_G$ oscillate and contain infinite zeros at infinite, therefore finite magnetic energy cannot be realized \cite{bogoyavlenskij2000astrophysical}. For the separated solution (\ref{eq:Sep_solution_2}), after choosing some $\psi_0 = \const$ to be the boundary of the plasma domain and utilizing (\ref{eq:Boundary_Condition}) for the current sheet, finite magnetic energy can be achieved for this type of pressure profile.

\subsubsection{Examples of Second family of axial solutions}
\begin{figure}[htbp]	
\begin{center}
\includegraphics[width = 1\textwidth]{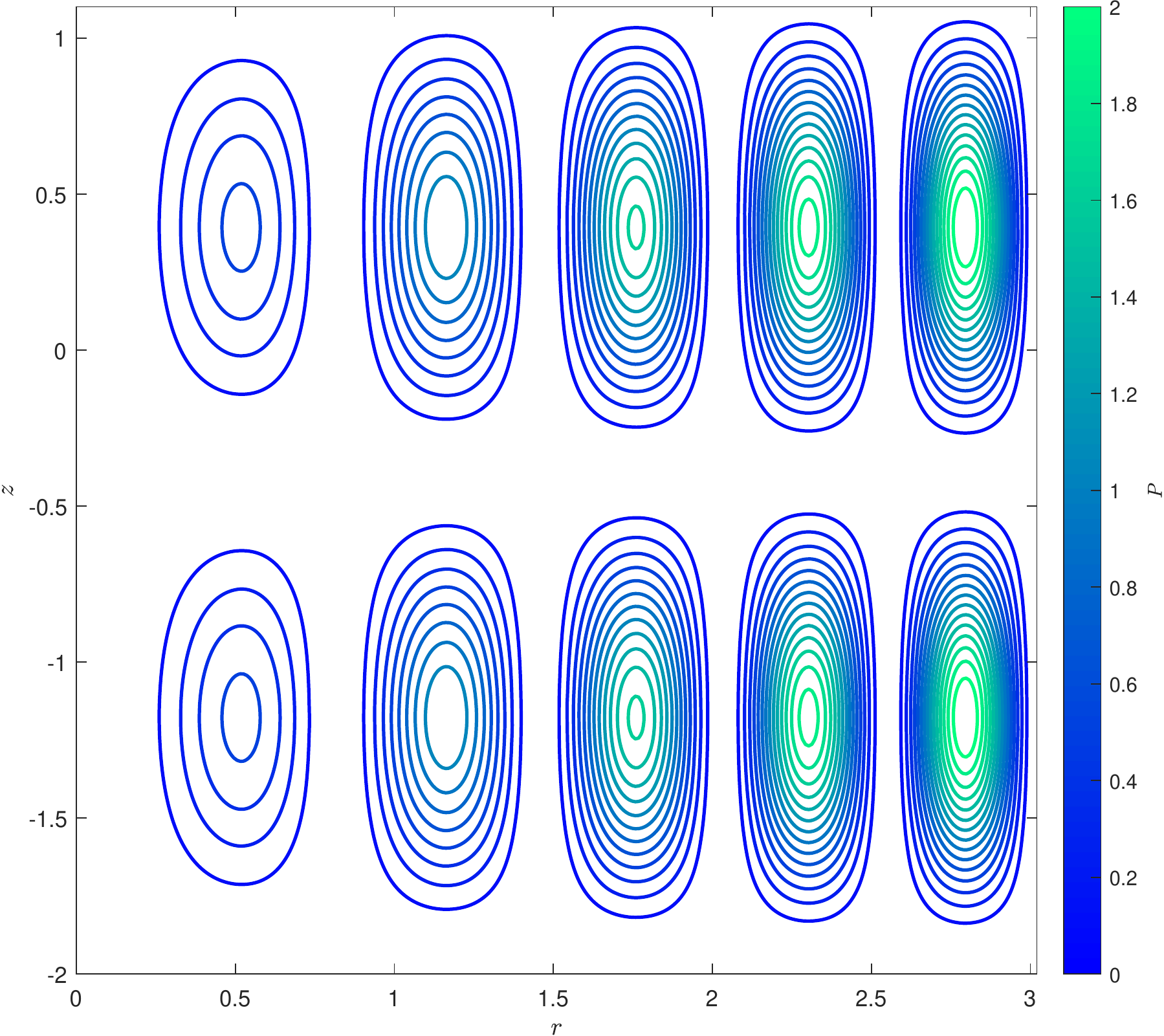}
\end{center}
\caption{\label{fig:Axial_family2_pressurecontour_global}A cross-section of magnetic surfaces $\psi, P = \const$ for a sample axially symmetric plasma equilibrium solution belonging to a Family 2, \eqref{Sep_solution_2}, with $C_1 = 1$, $C_2 = 0$, $C_3 = 1$, $C_4 = 1$, $k = 2$, $\alpha = 5$, and $q = \sqrt{3}$. The picture is periodic in $r$. The colorbar shows the values of the dimensionless pressure $P = P_0 + q^2\frac{\psi^2}{2}$}
\end{figure}

\begin{figure}[h]
	\centering
	\begin{subfigure}{0.32\textwidth}
		\centering
		\includegraphics[width=\textwidth]{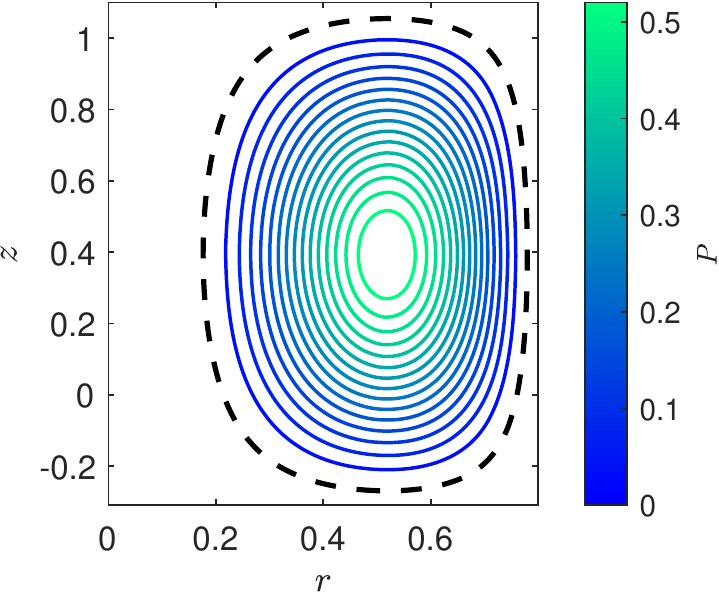}
		\caption{}\label{fig:Axial_Coulomb_contour_pressure}
	\end{subfigure}
	\hfill
	\begin{subfigure}{0.32\textwidth}
		\centering
		\includegraphics[width=\textwidth]{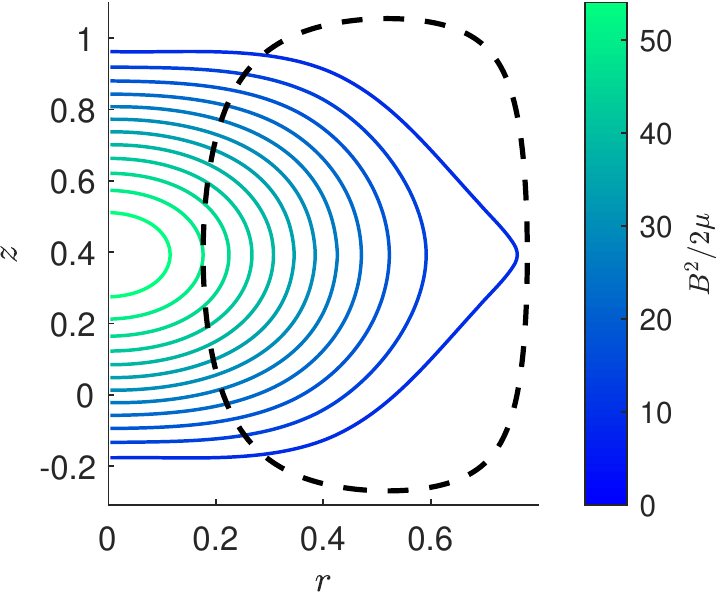}
		\caption{}\label{fig:Axial_family2_mag_energy_density}
	\end{subfigure}
	\hfill
	\begin{subfigure}{0.32\textwidth}
		\centering
		\includegraphics[width=\textwidth]{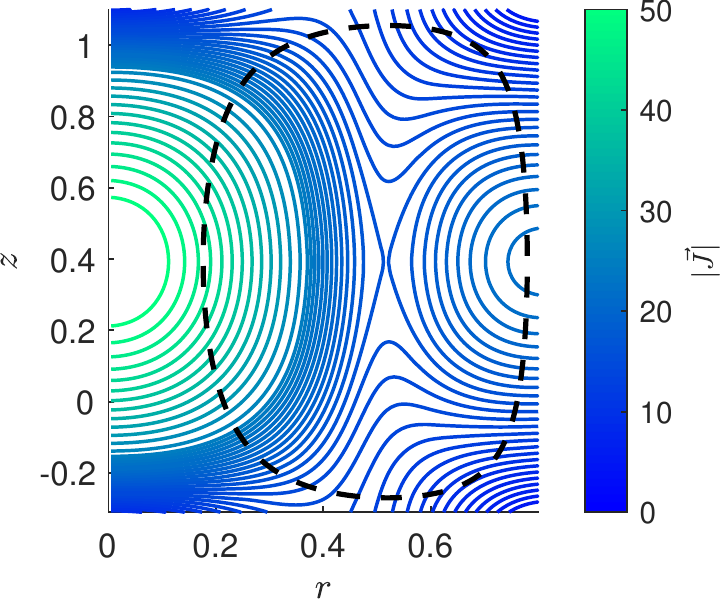}
		\caption{}
		\label{fig:Axial_family2_current_density}
	\end{subfigure}
	\caption{A cross-section of magnetic surfaces $\psi, P = \const$ for a sample axially symmetric plasma equilibrium solution belonging to a Family 2, \eqref{Sep_solution_2}, with $C_1 = 1$, $C_2 = 0$, $C_3 = 1$, $C_4 = 1$, $k = 2$, $\alpha = 5$, and $q = \sqrt{3}$. The picture is periodic in $r$. The colorbar shows the values of the dimensionless pressure $P = P_0 + q^2\frac{\psi^2}{2}$. The truncated boundary is shown boldface which coincides with the current sheet marking the boundary of the plasma.Lines of constant}\label{fig:axial_family2}
\end{figure}

\begin{figure}[htbp]
\begin{center}
\includegraphics[width=0.5\textwidth]{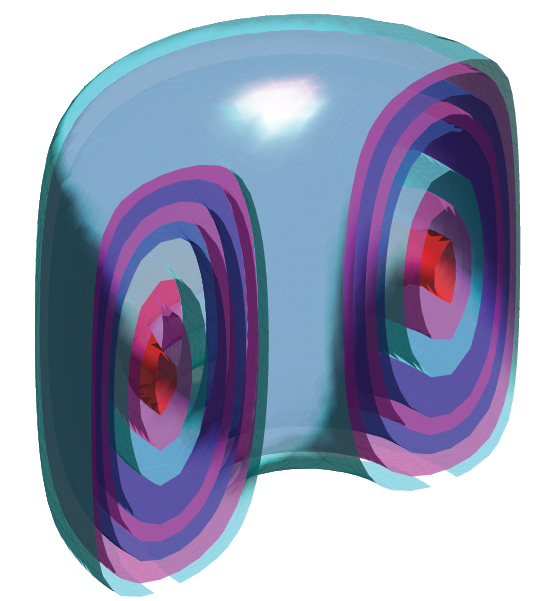}
\end{center}
\caption{\label{fig:Axial_family2_surf}Axially symmetric magnetic surfaces $\psi, P=\const$ for Family 2, \eqref{Sep_solution_2}, with $C_1 = 1$, $C_2 = 0$, $C_3 = 1$, $C_4 = 1$, $k = 2$, $\alpha = 5$, and $q = \sqrt{3}$ shown in 3D by rotating \Figref{Axial_Coulomb_contour_pressure} about the z axis.}
\end{figure}

Utilizing \eqref{Sep_solution_2}, the contours of the pressure profile can be constructed as seen in \Figref{Axial_family2_pressurecontour_global}. However, this solution has infinitely many zeroes as $r \to \infty$ and will have infinite magnetic energy in any layer of $z$ \cite{bogoyavlenskij2000astrophysical}. Therefore, by choosing some pressure contour, $P_0$ , to mark the boundary of the plasma domain, outside of which the pressure and the magnetic field are set to zero, $P = \vec{B} = 0$, the \emph{truncated} solution will now contain finite magnetic energy inside of this domain. The truncated solution can be seen in \Figref{Axial_Coulomb_contour_pressure}. One should note the positive pressure profile inside of the chosen domain. The magnetic energy density inside of the domain is seen in \Figref{Axial_family2_mag_energy_density}. A 3D magnetic surface generated by rotating \Figref{Axial_Coulomb_contour_pressure} about the $z$ axis can then be seen in \Figref{Axial_family2_surf}.
\medskip

\begin{figure}[h]
	\centering
	\begin{subfigure}{0.48\textwidth}
		\centering
		\includegraphics[width=\textwidth]{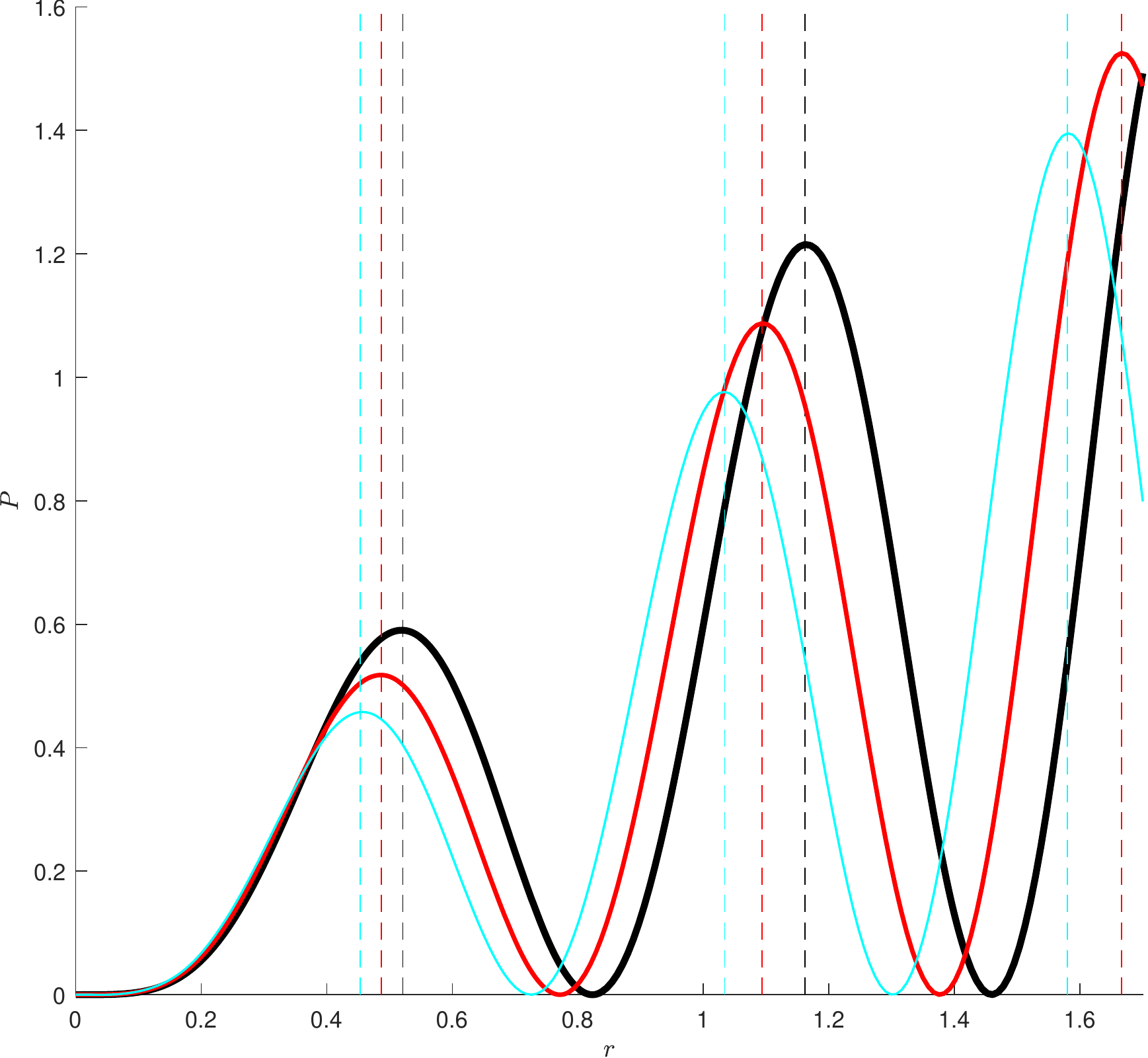}
		\caption{}\label{fig:Axial_family2_Mag_energy_density}
	\end{subfigure}
	\hfill
	\begin{subfigure}{0.48\textwidth}
		\hfill
		\centering
		\includegraphics[width=\textwidth]{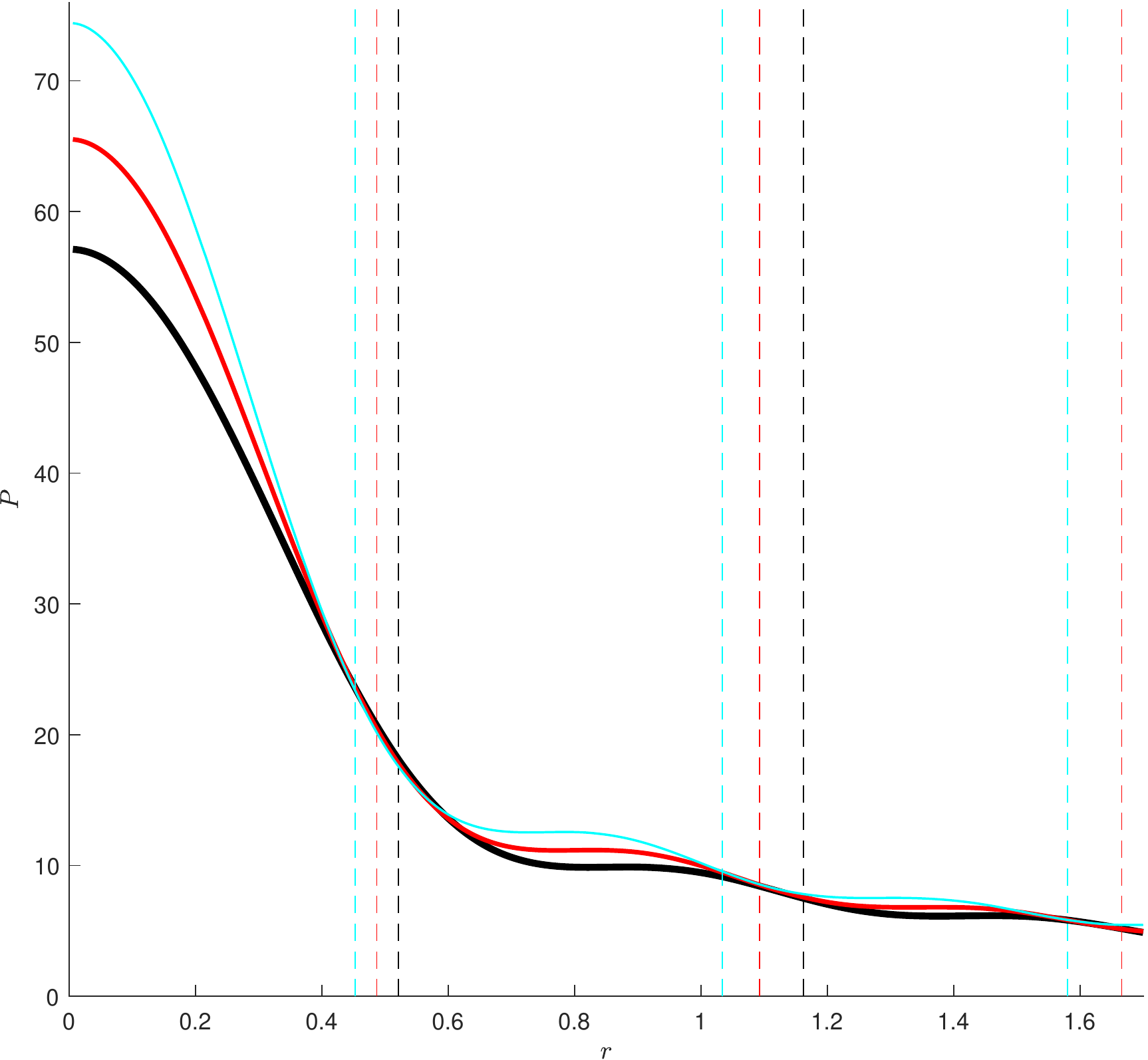}
		\caption{}
		\label{fig:Axial_family2_Mag_energy_density}
	\end{subfigure}
	\caption{A pressure plot of the solution from Figure \ref{fig:axial_fam1} for $z = 0$. The asymptotics of the magnetic energy densities can be seen in Figure \ref{fig:Axial_Whittaker_mag_energy_plot}.}\label{fig:Axial_fam1_plot}
\end{figure}

\section{Helically symmetric exact plasma equilibria and exact solutions}

As with the axial symmetry, similarly, the JFKO equation, \eqref{JFKO}, for certain choices of the arbitrary functions $P$ and $I$ becomes a linear PDE in which a separated solution can be sought. Again, choosing the pressure $P$ to be quadratic and the arbitrary function $I$ to be linear:
\begin{equation}\label{eq:P_I_helical_linear}
P(\psi) = P_0 + b\psi + \frac{1}{2}\sigma\psi^2, \quad
I(\psi) = \alpha \psi,
\end{equation}
the JFKO equation becomes a linear homogeneous second order PDE:
\begin{equation}
\frac{1}{r^2} \frac{\partial^2 \psi}{\partial \xi^2} + \frac{1}{r} \frac{\partial}{\partial r}\Big(\frac{r}{r^2 + \gamma^2}\frac{\partial \psi}{\partial r}\Big)  + \frac{\alpha^2 \psi}{r^2 + \gamma^2} + \frac{2\gamma \alpha \psi}{(r^2 + \gamma^2)^2} + 4\sigma \psi= -b,
\end{equation}
again the homogenous case, $b = 0$, is considered,

\begin{equation}\label{eq:JFKO_lin}
\frac{1}{r^2} \frac{\partial^2 \psi}{\partial \xi^2} + \frac{1}{r} \frac{\partial}{\partial r}\Big(\frac{r}{r^2 + \gamma^2}\frac{\partial \psi}{\partial r}\Big)  + \frac{\alpha^2 \psi}{r^2 + \gamma^2} + \frac{2\gamma \alpha \psi}{(r^2 + \gamma^2)^2} + 4\sigma \psi= 0,
\end{equation}

which admits separated solutions $\psi(r,u)= R(r)\Xi(\xi)$, satisfying
\begin{subequations}
\begin{equation}
\Xi'' = \lambda \Xi,
\end{equation}
\begin{equation}\label{eq:R_problem3}
r\Big(\frac{r}{r^2 + \gamma^2}R'\Big)' + \Big(\frac{\alpha^2r^2}{r^2 + \gamma^2} + \frac{2\gamma \alpha r^2}{(r^2 + \gamma^2)^2} + 4ar^2\Big)R = -\lambda R.
\end{equation}
\end{subequations}
Here the separation constant, $\lambda$ is taken to be negative $\lambda = -\omega^2$  which, as with the axial case, corresponds to a model of a plasma jet stretched along the $z$ axis. Therefore, $\Xi$ will have solutions in terms of sines and cosines.
Depending on the value of $\sigma$ in the pressure term of (\ref{eq:P_I_helical_linear}), one again obtains two different families of solutions. The two families of solutions correspond to two different types of pressure profiles. For $\sigma<0$, $P\leq P_1=\const>0$ with $P\to P_1$ when $|\vec{x}|\to\infty$. For the case when $\sigma>0$, $P>0$ inside of the plasma domain $\mathcal{V}$ and  $P=0$ outside of $\mathcal{V}$ similar to the axial cases discussed previously.

\subsection{The first family of helically symmetric solutions}\label{helical}

The first family of solutions arises when $\sigma=-\kappa^2<0$. This corresponds to a pressure profile which is bounded above by some $P_1$ where $P \to P_1$ when $|\vec{x}| \to \infty$. This model is more appropriate for plasmas residing in atmosphere. Upon the substitution of this pressure form and $\lambda = -\omega^2$ the following $R(r)$ equation arises:
\begin{equation}\label{eq:R_problem2_exact1}
r\Big(\frac{r}{r^2 + \gamma^2}R'\Big)' + \Big(\frac{\alpha^2r^2}{r^2 + \gamma^2} + \frac{2\gamma \alpha r^2}{(r^2 + \gamma^2)^2} + 4\kappa^2r^2\Big)R = \omega^2 R.
\end{equation}
and the $\Xi$ equation:
\begin{equation*}
\Xi'' = -\omega^2 \Xi.
\end{equation*}
which has the solution:
\begin{equation}\label{eq:Xi_solution}
\Xi(\xi) = C_1\sin(\omega \xi) + C_2 \cos(\omega \xi).
\end{equation}
The solution to the $r$ differential equation \eqref{R_problem2_exact1} can be written in terms of the confluent Heun function as:
\begin{equation}\label{eq:R_solution1_JFKO}
R(r) = e^{-\kappa r^2}\Big(C_1r^b \mathcal{H}_C(a,b,-2,c,d,-r^2/\gamma^2) + C_2 r^{-b}\mathcal{H}_C(a,-b,-2,c,d,-r^2/\gamma^2)\Big).
\end{equation}
Here
\begin{subequations}\label{eq:Heun_Parameters}
\begin{equation}
a = \kappa \gamma^2,
\end{equation}
\begin{equation}
b = \gamma \omega,
\end{equation}
\begin{equation}
c = \frac{\gamma^2(\gamma^2\kappa^2 - \alpha^2 + \omega^2)}{4},
\end{equation}
\begin{equation}
d = 1 - \frac{\kappa^2 \gamma^2}{4} + \frac{\alpha^2 - \omega^2}{4}\gamma^2 + \frac{\alpha
\gamma}{2},
\end{equation}
\end{subequations}
Where the confluent Heun function $\mathcal{H}_C(a,b,-2,c,d,x)$ satisfies the following confluent Heun ODE:
\begin{equation}\label{eq:Confluent_Heun_ODE}
y'' - \dfrac{(-x^2 a + (-b + a)x + b + 1)}{x(x-1)}y' - \dfrac{((-ba -2c)x + (b+1)a + b - 2d + 2)}{2x(x-1)} y = 0.
\end{equation}
Some important remarks about the solution to \eqref{R_problem2_exact1} and the confluent Heun function are:
\begin{enumerate}\label{Heun Remarks}
\item
For the following examples shown, $C_2$ will be set to zero.

\item
There exists necessary and sufficient conditions for the confluent Heun function to produce polynomials which are discussed thoroughly in \cite{saad2015solvability}. These solutions correspond with solution discussed in \cite{bogoyavlenskij2000helically}.
\end{enumerate}
\medskip

Considering the confluent Heun function $\mathcal{H}_C(\alpha, \beta, -2, \delta, \eta, x)$, as a solution to \eqref{Confluent_Heun_ODE}. A necessary condition for the emergence of these polynomials is $\delta = -\alpha(n + \beta/2)$ where n is some positive integer which specifies the degree of this polynomial. The needed sufficient condition for the polynomials of this function come from choosing a finite number of characteristic values for $\eta$. These characteristic values are chosen such to be the roots of the coefficient of the $(n +1)$ degree of the series expansion. Further details on the conditions for these polynomials can be found in \cite{saad2015solvability}.
\medskip

Due to the first remark, the second solution to \eqref{R_problem2_exact1} in which the second parameter of $\mathcal{H}_C$ is a negative value is removed for the use in this paper. Therefore a separated solution for this first family of helically symmetric solutions can then be written as:
\begin{equation}\label{eq:Sep_solution_3}
\psi_\omega(r,\xi) = e^{-\kappa r^2}r^b \mathcal{H}_C(a,b,-2,c,d,{-r^2}/{\gamma^2})(C_1\sin(\omega \xi) +  C_2\cos(\omega \xi)).
\end{equation}

A general solution can be written as any linear combination of \eqref{Sep_solution_3} by the following:
\begin{equation}\label{eq: General Solution3}
  \psi(r,\xi) = \int_{-\infty}^{\infty}\psi_\omega(r,\xi)\;d \omega.
\end{equation}
assuming $C_i = C_i(\omega)$, $i = 1,2$.
\subsection{Examples of First family of helical solutions}
Similar to the axial case, with the first family there are two separate cases to consider. The first is the most general solution of this pressure type written in terms of the confluent Heun function. This separated solution, \eqref{Sep_solution_3} has a special case in which the confluent Heun function produces polynomials, this is the second case to consider and these solutions are the same as the solution discussed in \cite{bogoyavlenskij2000helically} which has the general solution \eqref{helical_jets} which is written as a linear combination of all polynomial solutions.
\medskip

\begin{figure}[htb!]
	\centering
	\begin{subfigure}{0.32\textwidth}
		\centering
		\includegraphics[width=\textwidth]{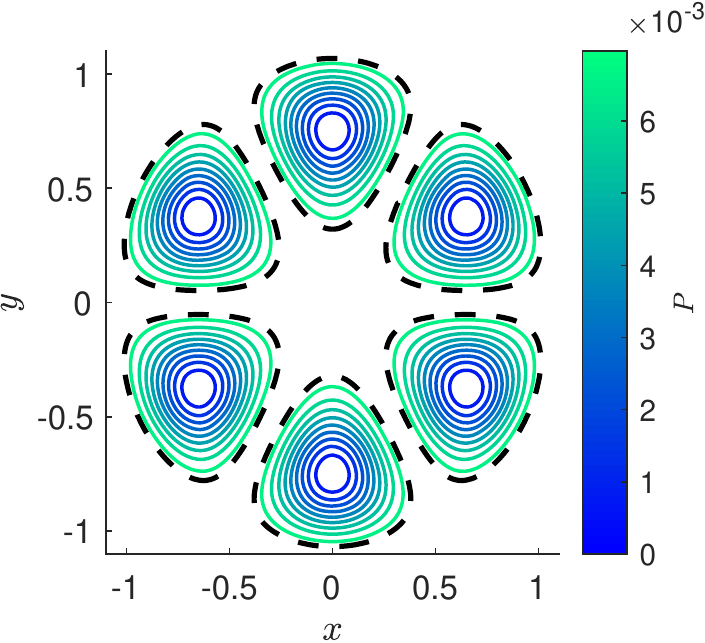}
		\caption{}\label{fig:Helical_family1_pressure_contour_xy}
	\end{subfigure}
	\hfill
	\begin{subfigure}{0.32\textwidth}
		\centering
		\includegraphics[width=\textwidth]{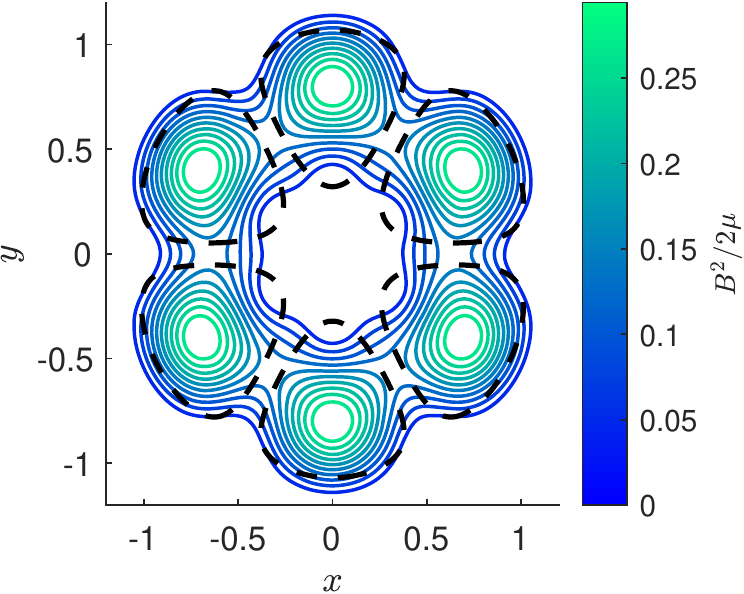}
		\caption{}\label{fig:Helical_family1_mag_energy_contour_xy}
	\end{subfigure}
	\hfill
	\begin{subfigure}{0.32\textwidth}
		\centering
		\includegraphics[width=\textwidth]{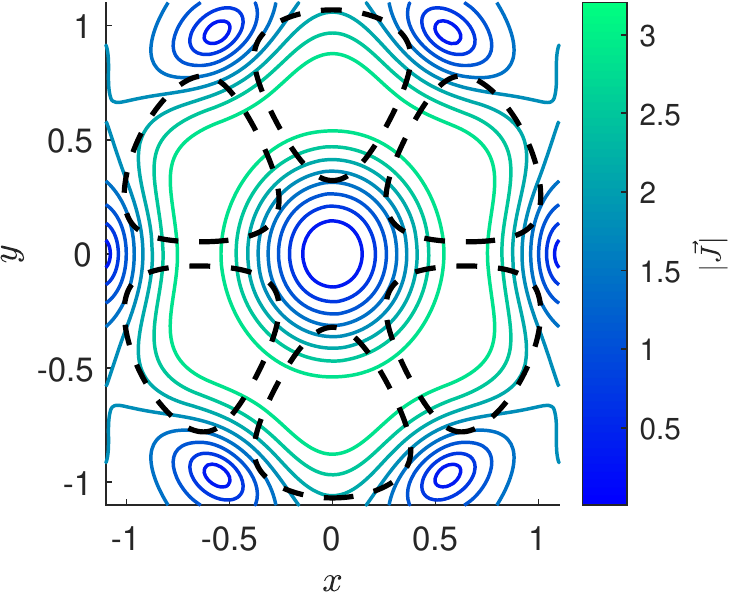}
		\caption{}
		\label{fig:Helical_family1_current_density_contour_xy}
	\end{subfigure}
	\caption{A truncated helically symmetric physical solution. The pressure contour, magnetic energy density and current density magnitude can be seen from left to right.}\label{fig:Helical_family1}
\end{figure}

Unlike the axial case for which only the physical global solutions were the polynomial solutions written by \eqref{axial_jets}, there exists global solutions in which the confluent Heun function does not produce polynomials. An example of this more general type of solution given by \eqref{Sep_solution_3} with the parameters given by (\ref{eq:Heun_Parameters}) can be seen in \Figref{Helical_family1_pressure_contour_xy} with the corresponding magnetic energy density found in \figref{Helical_family1_mag_energy_contour_xy}.
\medskip

The second example from this family of solutions is for the case when the confluent Heun function produces polynomials. This is the same solution as the ones discussed in \cite{bogoyavlenskij2000helically} by Bogoyavlenkij. Therefore, for this special case, a solution of that type was utilized with the solution given by \eqref{helical_jets}. The constant pressure contour can be seen by \Figref{Helical_family1_pressure_contour_xy} and the corresponding magnetic energy density in \Figref{Helical_family1_mag_energy_contour_xy}.
\medskip

With a lifting and rotating motion along and about the $z$ axis, the magnetic surfaces in which both $\vec{B}$ and $\curl \vec{B}$ are tangent can be seen in 

\begin{figure}[htb!]
	\begin{center}
		\includegraphics[width=.7\textwidth]{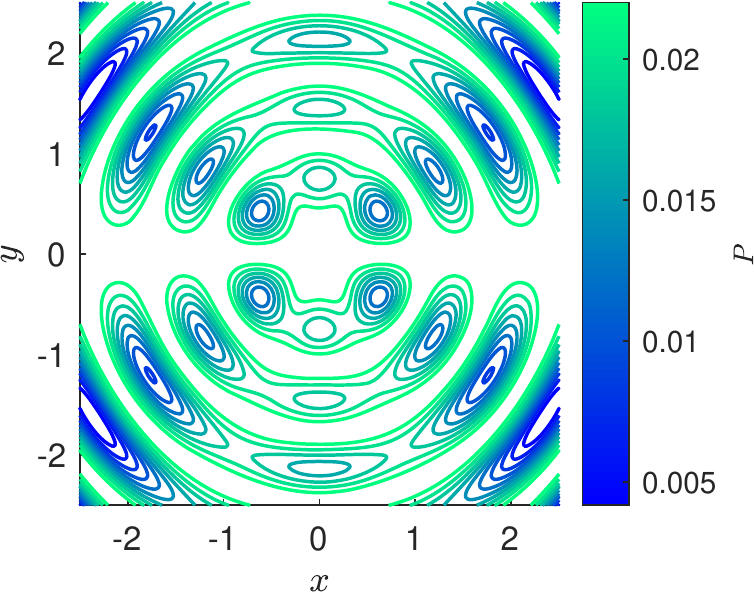}
		\caption{Pressure profile for a linear combination of  $\psi$ for Family 1 where $c \neq -a(n + (b/2))$. Here $\Psi(r,\xi) = \psi_1(r,\xi) + \psi_2(r,\xi)$ where $\psi_1(r,\xi)$ is given in the previous figure and $\psi_2(r,\xi))$ is given by (\ref{eq:Sep_solution_3}) with $\alpha = 5.9$, $\kappa = 1$, $\gamma = 1$, $\omega = 2$, $C_1 = 1$ and $C_2 = 0$. Here the solution is non-physical as it grows unbounded unless one restricts the plasma domain to within one helical cylinder or so.}\label{fig:Helical_family1_lincomb}
	\end{center}
\end{figure}

An interesting linear combination of a non-physical solution can be seen in Figure \ref{fig:Helical_family1_lincomb}.

\begin{figure}[htbp]
\begin{center}
\includegraphics[width=1\textwidth]{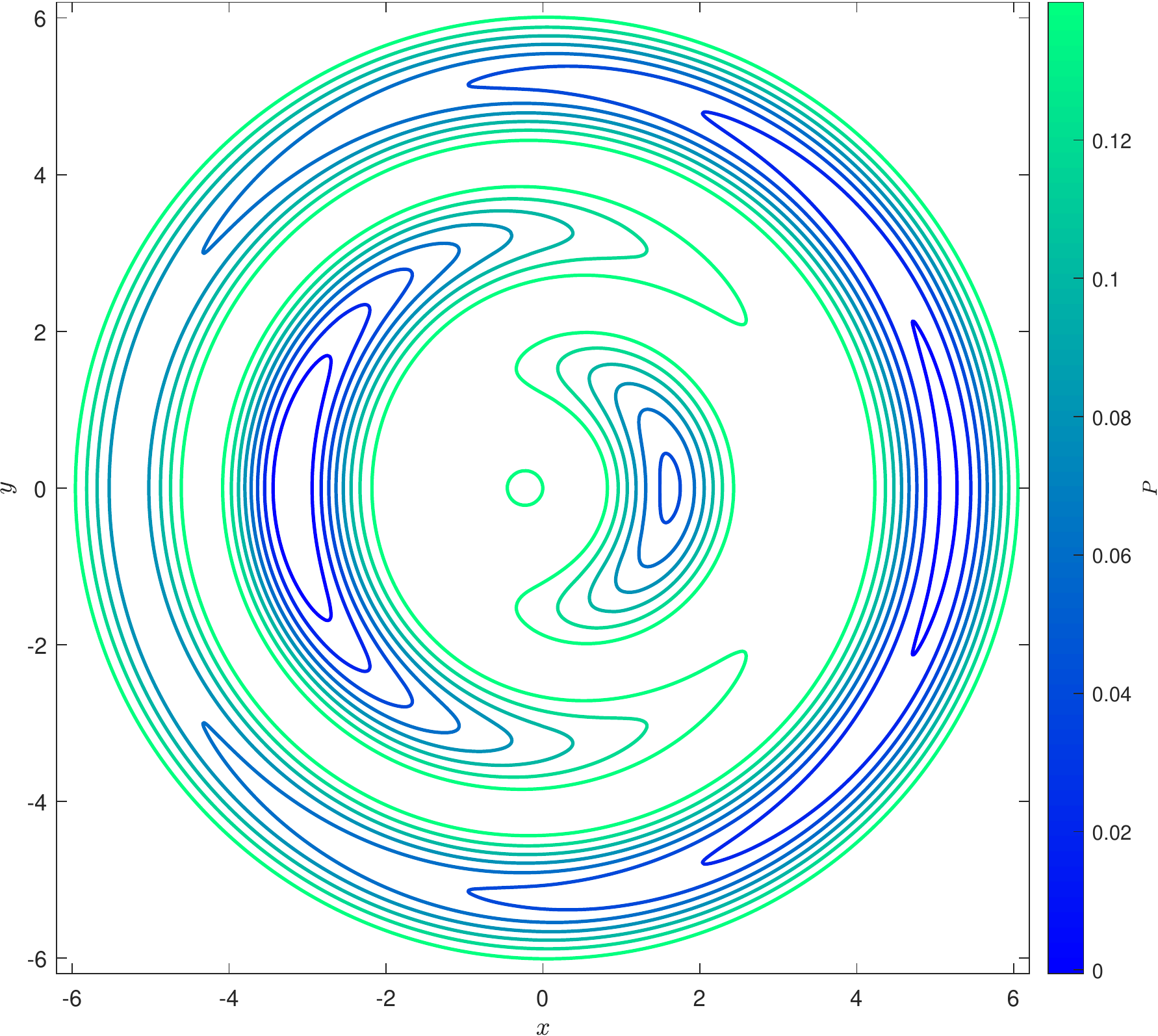}
\caption{\label{fig:Helical_bog_contour_pressure}A cross-section of magnetic surfaces $\psi, P = \const$ for a sample helically symmetric plasma equilibrium solution belonging to a special case of Family 1, \eqref{helical_jets}, for $N=4$, $n = 0$, $m = 1$, $\beta=0.1$, $aN = an = 1$ and $bn = 0$. The colorbar shows the values of the dimensionless pressure, $P = P_0 - 2\beta^2\psi^2$, $P_0 = 1$.}
\end{center}
\end{figure}

\begin{figure}[htbp]
\begin{center}
\includegraphics[width=1\textwidth]{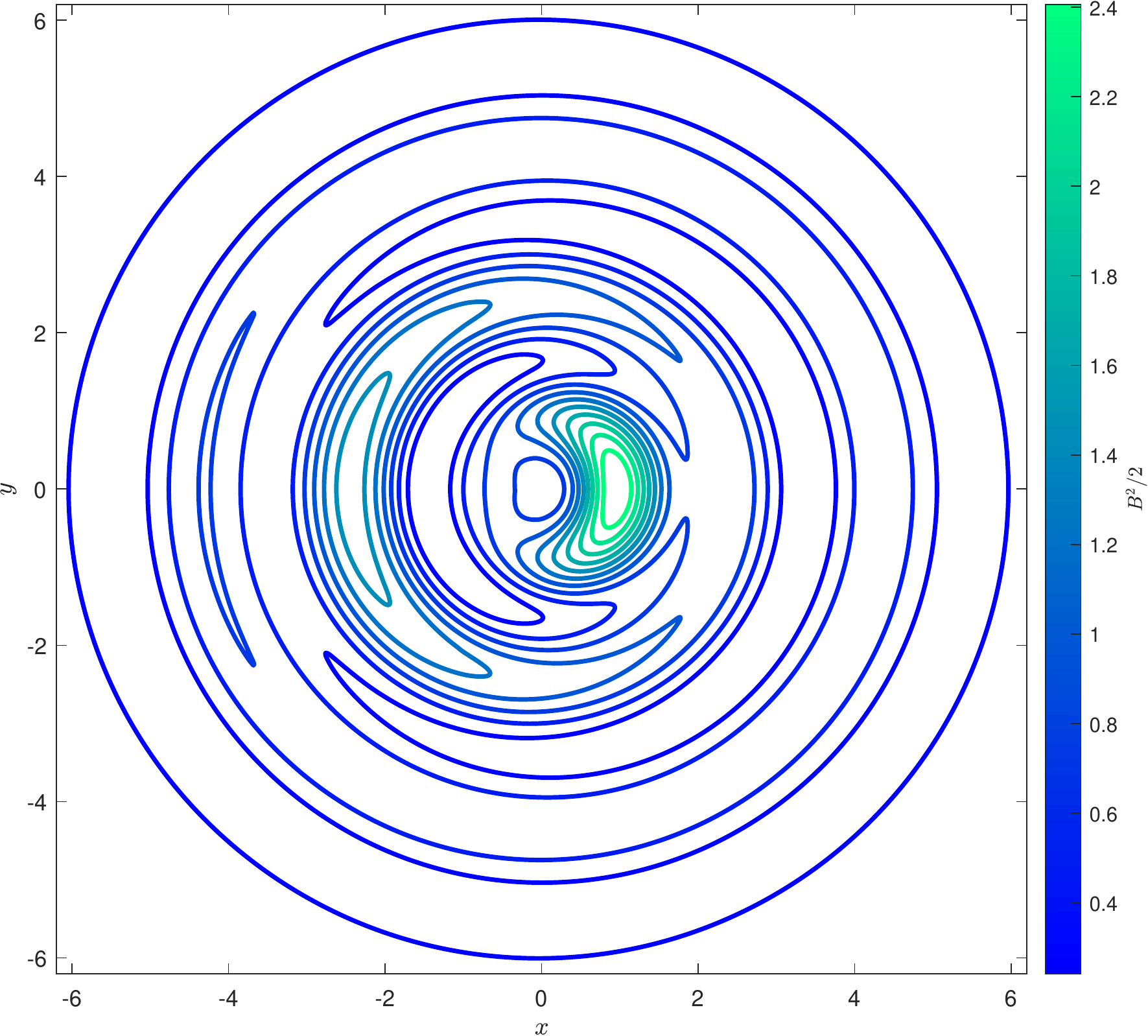}
\caption{\label{fig:Helical_bog_mag_energy_density}A cross-section of magnetic energy density for a sample helically symmetric plasma equilibrium solution belonging to a special case of Family 1, \eqref{helical_jets}, for $N=4$, $n = 0$, $m = 1$, $\beta=0.1$, $a_N = an = 1$ and $bn = 0$. The colorbar shows the values of the dimensionless magnetic energy density.}
\end{center}
\end{figure}

\begin{figure}[htbp]
\begin{center}
\includegraphics[width=1\textwidth]{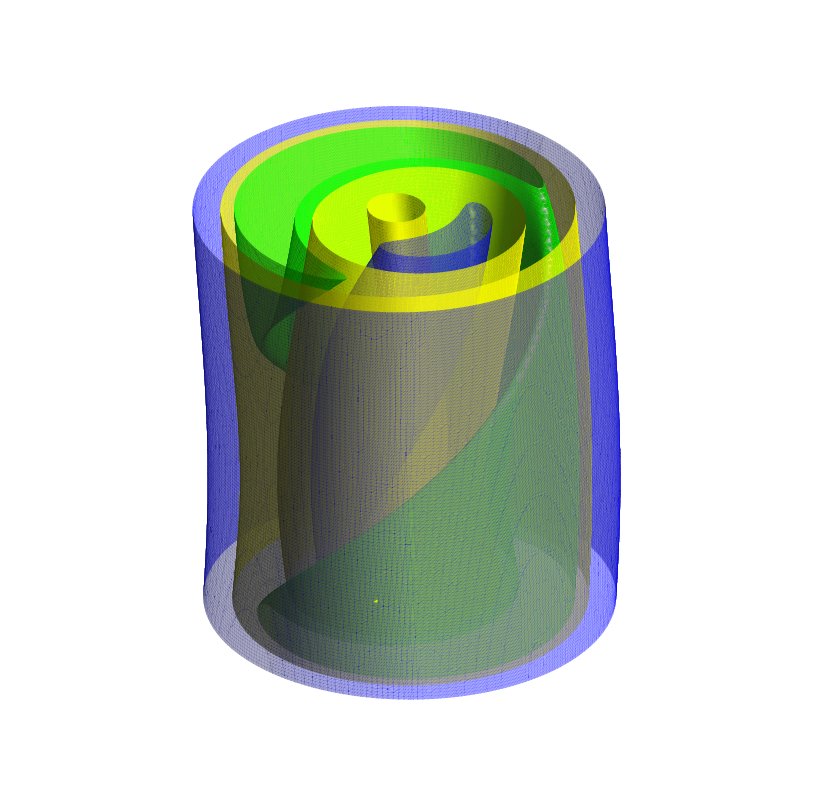}
\caption{\label{fig:Helical_bog_surf}Helically symmetric magnetic surfaces $\psi, P=\const$ for special case of Family 1, \eqref{axial_jets}, with $N=4$, $n = 0$, $m = 1$, $\beta=0.1$, $a_N = an = 1$ and $bn = 0$.}
\end{center}
\end{figure}

\subsection{The second family of helically symmetric solutions}
The second family arises when $\sigma = \kappa^2 > 0$ in \eqref{JFKO_lin}. This corresponds to the pressure profile in which $P>0$ inside of $\mathcal{V}$ and $P=0$ outside suitable for plasma in vacuum. After substituting this form of pressure into \eqref{R_problem2}, the following ODE similar to \eqref{R_problem2_exact1} can be seen:
\begin{equation}\label{eq:R_problem2_exact2}
r\Big(\frac{r}{r^2 + \gamma^2}R'\Big)' + \Big(\frac{\alpha^2r^2}{r^2 + \gamma^2} + \frac{2\gamma \alpha r^2}{(r^2 + \gamma^2)^2} - 4\kappa^2r^2\Big)R = \omega^2 R.
\end{equation}
Along with the same $\Xi$ equation given above with the solution given by \eqref{Xi_solution}. The solution to \ref{eq:R_problem2_exact2} can again be written in terms of the confluent Heun function, $\mathcal{H}_C$, this time with a complex exponential and complex parameters.
\begin{equation}\label{eq:R_solution2_JFKO}
  R(r) = C_1 r^{b} e^{(-\frac{i}{2} \kappa r^2)}H_C(ia,b,-2,c,d,\frac{-r^2}{\gamma^2}) + C_2 r^{b} e^{(-\frac{i}{2} \kappa r^2)}H_C(ia,-b,-2,c,d,\frac{-r^2}{\gamma^2}).
\end{equation}
Where $a$, $b$, $c$, and $d$ are given above in \eqref{Heun_Parameters}.
\medskip

Following the 2nd remark about the confluent Heun function from section \ref{Heun Remarks}, we set $C_2 = 0$ in (\ref{eq:R_solution2_JFKO}). It should be noted, as it is not obvious, that the first solution of \eqref{R_solution2_JFKO} is a real valued function for real $a$, $b$, $c$, $d$ and $r$. Therefore, a separated solution to \eqref{JFKO_lin} corresponding to helically symmetric plasma in vacuum is

\begin{equation}\label{eq:Sep_solution_4}
\psi_\omega(r,\xi) = e^{- i\kappa r^2} r^b \mathcal{H}_C(ia,b,-2,c,d,{-r^2}/{\gamma^2}) \left( C_1\sin(\omega \xi) + C_2\cos(\omega \xi)\right),
\end{equation}

Now a general solution can be written as

\begin{equation}\label{eq: General Solution4}
  \Psi(r,\xi) = \int_{-\infty}^{\infty}\psi_\omega(r,\xi)\;d \omega,
\end{equation}

where $C_1 = C_1(\omega)$, $C_2 = C_2(\omega)$ are arbitrary weighting distributions. The radial part, $R(r)$,  in the separated solution (\ref{eq:Sep_solution_4}) behaves periodically, similarity to the Coulomb wave functions described previously in section (3.2). Due to this, (\ref{eq:Sep_solution_4}) must be truncated at some chosen magnetic surface $\psi = \psi_0$ outside of which $P$ and $\vec{B} = 0$. This is again accomplished by utilizing the boundary condition (\ref{eq:Boundary_Condition}).

\subsection{An example of second family of helical solutions}
In the following example, the parameter $\gamma$ of the helical coordinate system (\ref{eq:helical_coordinates}) is set to $\gamma = 1$.
\medskip

\begin{figure}[htbp]	
\begin{center}
\includegraphics[width = 1\textwidth]{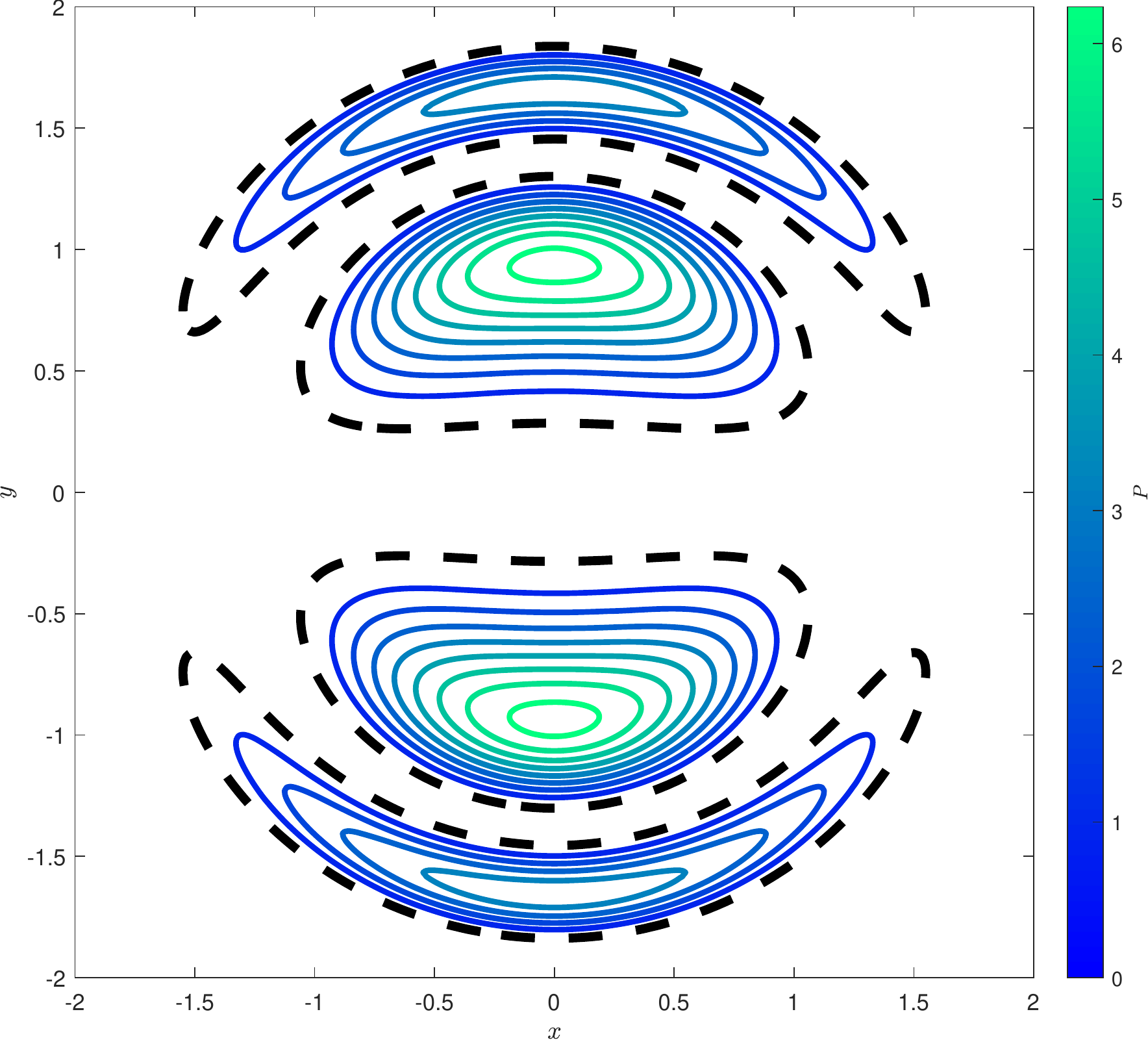}
\caption{\label{fig:Helical_family2_pressure_contour}A truncated cross-section of magnetic surfaces $\psi, P = \const$ for a sample helically symmetric plasma equilibrium solution belonging to Family 2 using \eqref{Sep_solution_4}, for $C_1 = 1$, $C_2 = 0$, $\alpha = 3$, $\kappa = 4$, $\gamma = 1$ and $\omega = 1$. The colorbar shows the values of the dimensionless pressure $P = \kappa^2\frac{\psi^2}{2}$. The black dashed border is the current sheet acting as the boundary of the plasma domain.}
\end{center}
\end{figure}

Using \eqref{Sep_solution_4} and choosing the following values for $\omega$, $C_1$, $C_2$, $\kappa$, $\gamma$ and $\alpha$ the following global contours of the pressure $P(\psi) = \frac{1}{2}\kappa^2\psi^2$ can be seen in \Figref{Helical_family2_pressure_contour}. It is important to note the periodic behaviour similar to the Coulomb wave functions seen in \Figref{Axial_family2_pressurecontour_global}. It appears that for the type of pressure configuration suitable for a plasma residing in vacuum the solutions for both axial symmetry and helical symmetry have oscillatory nature in the radial variable.
\medskip

\begin{figure}[htbp]	
\begin{center}
\includegraphics[width = 1\textwidth]{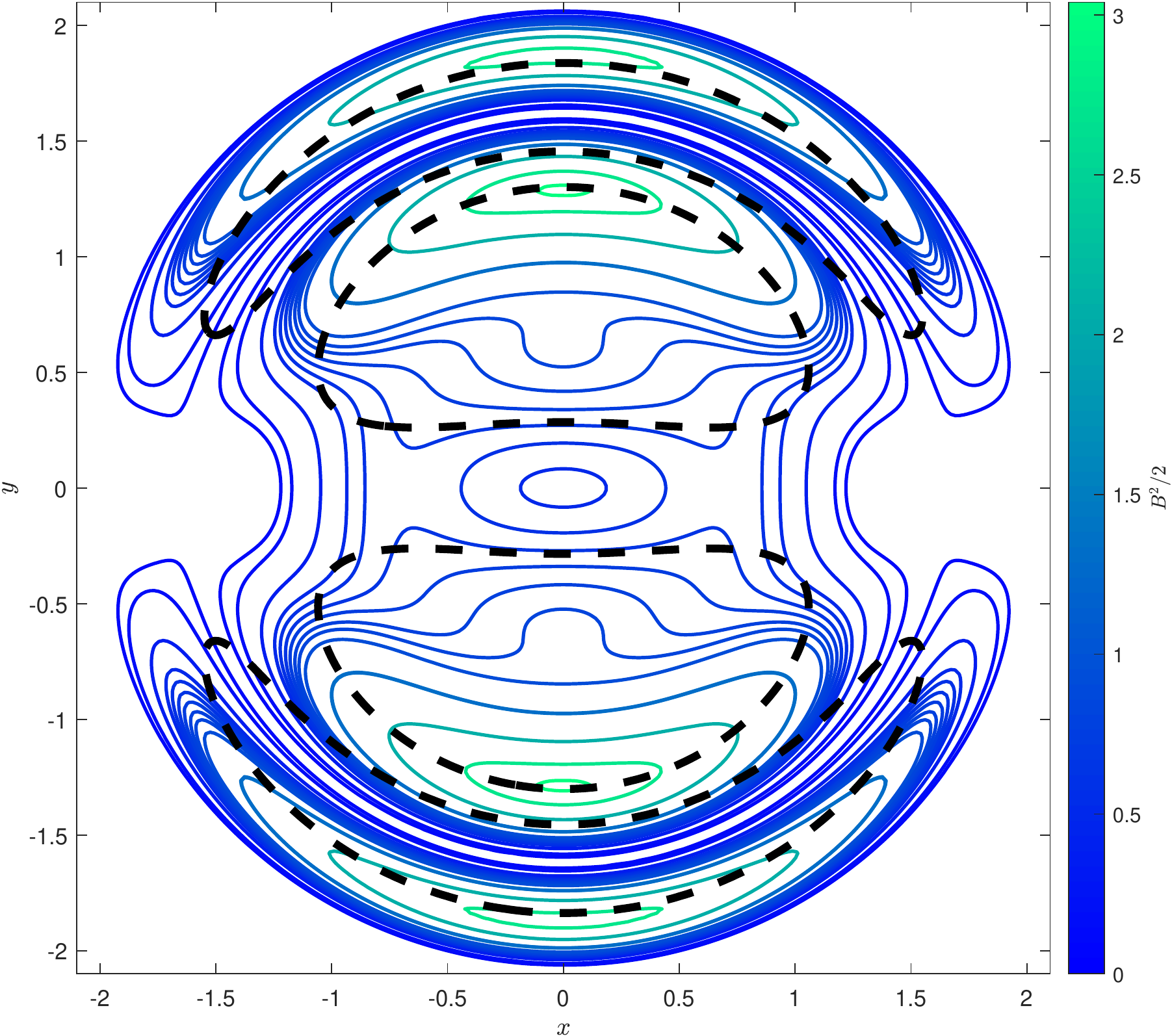}
\caption{\label{fig:Helical_family2_magnetic_energy} A cross-section of the magnetic energy density for a sample helically symmetric plasma equilibrium solution belonging to Family 2 using \eqref{Sep_solution_4} with $C_1 = 1$, $C_2 = 0$, $\alpha = 3$, $\kappa = 4$, $\gamma = 1$ and $\omega = 1$. Here the dotted bold black line corresponds to the boundary of the domain coinciding with the boldface border found in \Figref{Helical_family2_pressure_contour}.}
\end{center}
\end{figure}

\begin{figure}[htbp]	
\begin{center}
\includegraphics[width = .5\textwidth]{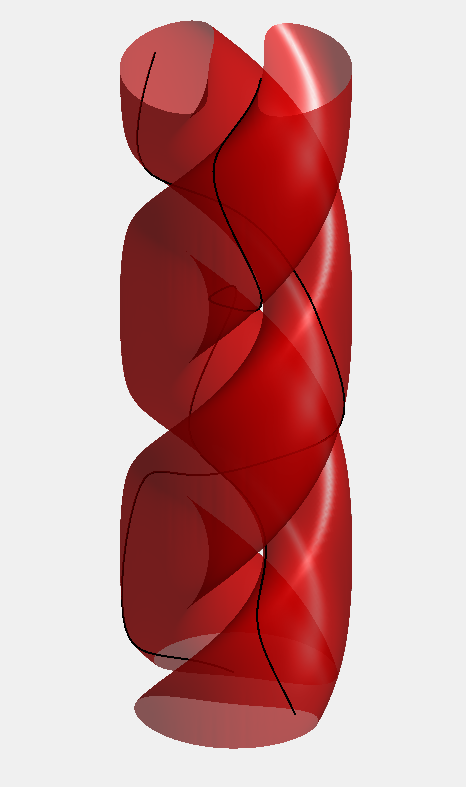}
\caption{\label{fig:Helical_family2_field_lines}A 3D magnetic surfaces $\psi, P = \const$ for a sample helically symmetric plasma equilibrium solution belonging to Family 2 using \eqref{Sep_solution_4} with $C_1 = 1$, $C_2 = 0$, $\alpha = 3$, $\kappa = 4$, $\gamma = 1$ and $\omega = 1$. Here the black lines represent the magnetic field lines tangent to this surface.}
\end{center}
\end{figure}

After truncating this solution at some chosen boundary $P(\psi) = P_0$, the pressure contour can be seen in \Figref{Helical_family2_pressure_contour} with the corresponding magnetic energy density seen in \Figref{Helical_family2_magnetic_energy}. For this example, magnetic field lines were also shown twisting up two separate nested helical surfaces as shown in \Figref{Helical_family2_field_lines}.

\begin{figure}[htbp]	
\begin{center}
\includegraphics[width = 1\textwidth]{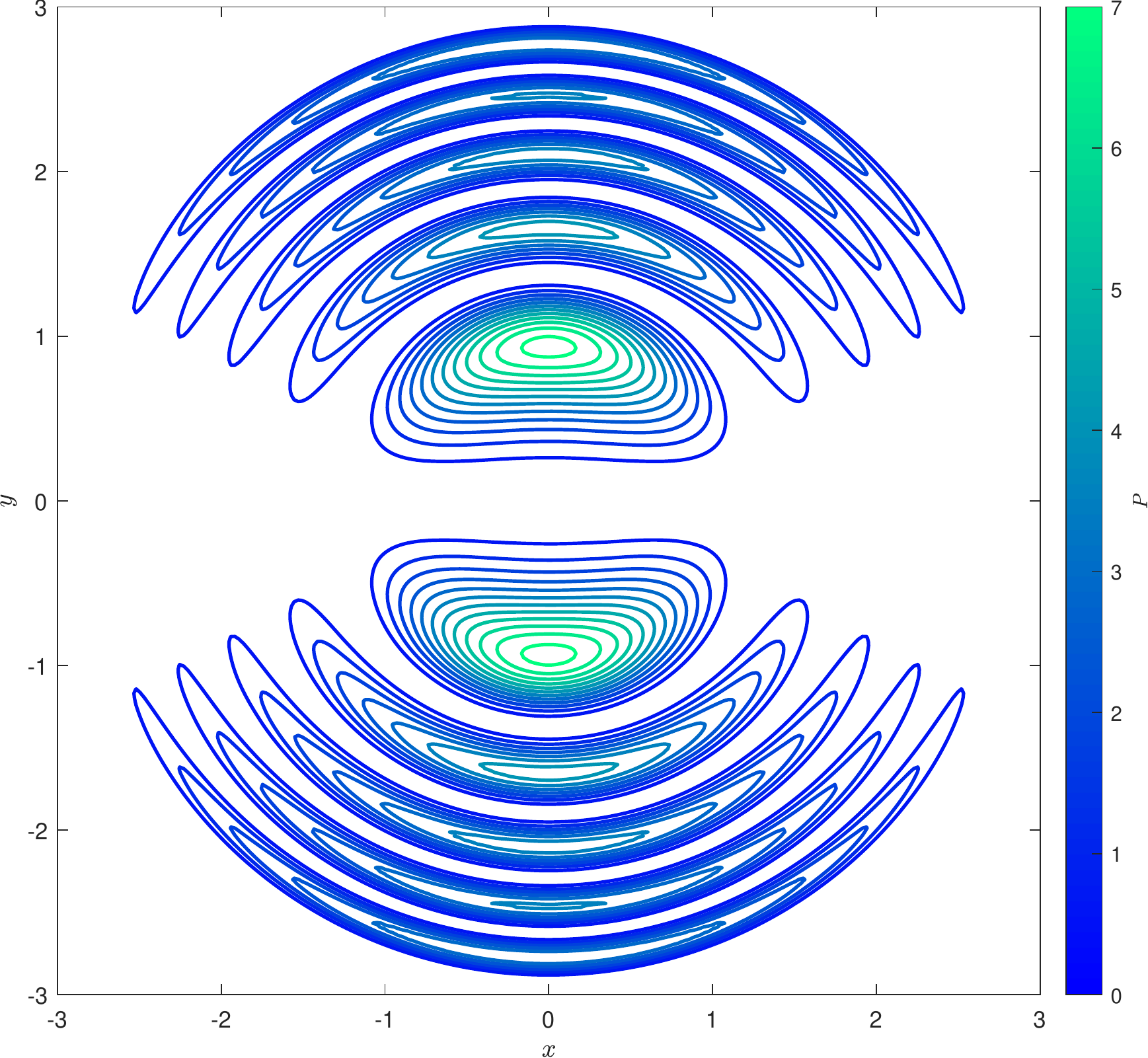}
\caption{\label{fig:Helical_family2_global}A cross-section of magnetic surfaces $\psi, P = \const$ for a sample helically symmetric plasma equilibrium solution belonging to Family 2 using \eqref{Sep_solution_4},  for $C_1 = 1$, $C_2 = 0$, $\alpha = 3$, $\kappa = 4$, $\gamma = 1$ and $\omega = 1$. The colorbar shows the values of the dimensionless pressure $P = 0.5\kappa^2\psi^2$.}
\end{center}
\end{figure}

\section{Generalization of the exact solution to the dynamic case}
As discussed in ref \cite{bogoyavlenskij2002symmetry}, the following families of intrinsic symmetries for \eqref{MHDeq} have the following form.
Letting $\vec{V}$, $\vec{B}$, $P$ and $\rho$ being a solution to \eqref{MHDeq} in which the density $\rho$ is constant on both magnetic field lines and streamlines. (that is $\rho = \rho(\psi)$), there exists an infinite family of solutions $\vec{V}_1$, $\vec{B}_1$, $P_1$ and $\rho_1$ which can be constructed by

\begin{subequations}\label{eq:Bog_trans}
\begin{equation}
\vec{B}_1 = b(\psi)\vec{B} + c(\psi)\sqrt{\mu \rho}\vec{V},
\end{equation}
\begin{equation}
\vec{V}_1 = \frac{c(\psi)}{a(\psi)\sqrt{\mu \rho}}\vec{B} + \frac{b(\psi)}{a(\psi)}\vec{V},
\end{equation}
\begin{equation}
P_1 = CP + \frac{C\vec{B}^2 - \vec{B}^2_1}{2\mu},
\end{equation}
\begin{equation}
\rho_1 = a^2(\psi)\rho;
\end{equation}
\end{subequations}
here, $a = a(\psi)$, $b = \b(\psi)$, $c = c(\psi)$ are constant on both magnetic fields lines and streamlines, with the only restriction, $b^2(\psi) - c^2(\psi) = C = \const$. The transformation (\ref{eq:Bog_trans}) is such as to preserve the magnetic surfaces of the initial plasma configuration $(\vec{V}, \vec{B}, P, \rho)$.\medskip

For the static equilibrium case ($\vec{V} = 0$), (\ref{eq:Bog_trans}) becomes

\begin{subequations}\label{eq:Bog_trans_2}
\begin{equation}
\vec{B}_1 = \sqrt{C + c^2(\psi)}\vec{B}, \quad \vec{V}_1 = \frac{c(\psi)}{a(\psi)\sqrt{\mu \rho}}\vec{B},
\end{equation}
\begin{equation}
P_1 = CP + -\frac{c^2(\psi)\vec{B}^2}{2\mu},
\end{equation}
\begin{equation}
\rho_1 = a^2(\psi)\rho.
\end{equation}
\end{subequations}

Written this way, it can be seen that for each solution of (\ref{eq:MHDst}) there exists a corresponding new solution depending on the freedom choice of two arbitrary functions, for example,

\begin{equation}\label{eq:trans_functions}
F_1(\psi) = c(\psi),\quad
F_2(\psi) = a(\psi)\sqrt{\rho(\psi)}.
\end{equation}

Therefore the transformations (\ref{eq:Bog_trans_2}) can then be written as

\begin{subequations}\label{eq:Bog_trans_st}
\begin{equation}
\vec{B}_1 = \sqrt{C + F_2^2(\psi)}\vec{B}, \quad \vec{V}_1 = \frac{F_2(\psi)}{F_1(\psi)\sqrt{\mu}}\vec{B},
\end{equation}
\begin{equation}
P_1 = CP -\frac{F_2^2(\psi)\vec{B}^2}{2\mu},
\end{equation}
\begin{equation}
\rho_1 = F_2^2(\psi),
\end{equation}
\end{subequations}

and used to transform the axially and helically symmetric exact solutions of sections (\ref{axial}) and (\ref{helical}) into new solutions with $\vec{V_1} \neq 0$.\medskip

It should be noted that while the original and transformed solutions may or may not be stable solutions of (\ref{eq:MHD}),Vladimirov in \cite{vladimirov1995general} showed that no new instability would arise from the transformations (\ref{eq:Bog_trans}). In other words, if one can show that the static equilibrium solutions are stable then these transformed solutions will also be stable.

\subsection{Transformation of axial family 2 example}

\begin{figure}[htbp]	
\begin{center}
\includegraphics[width = 1\textwidth]{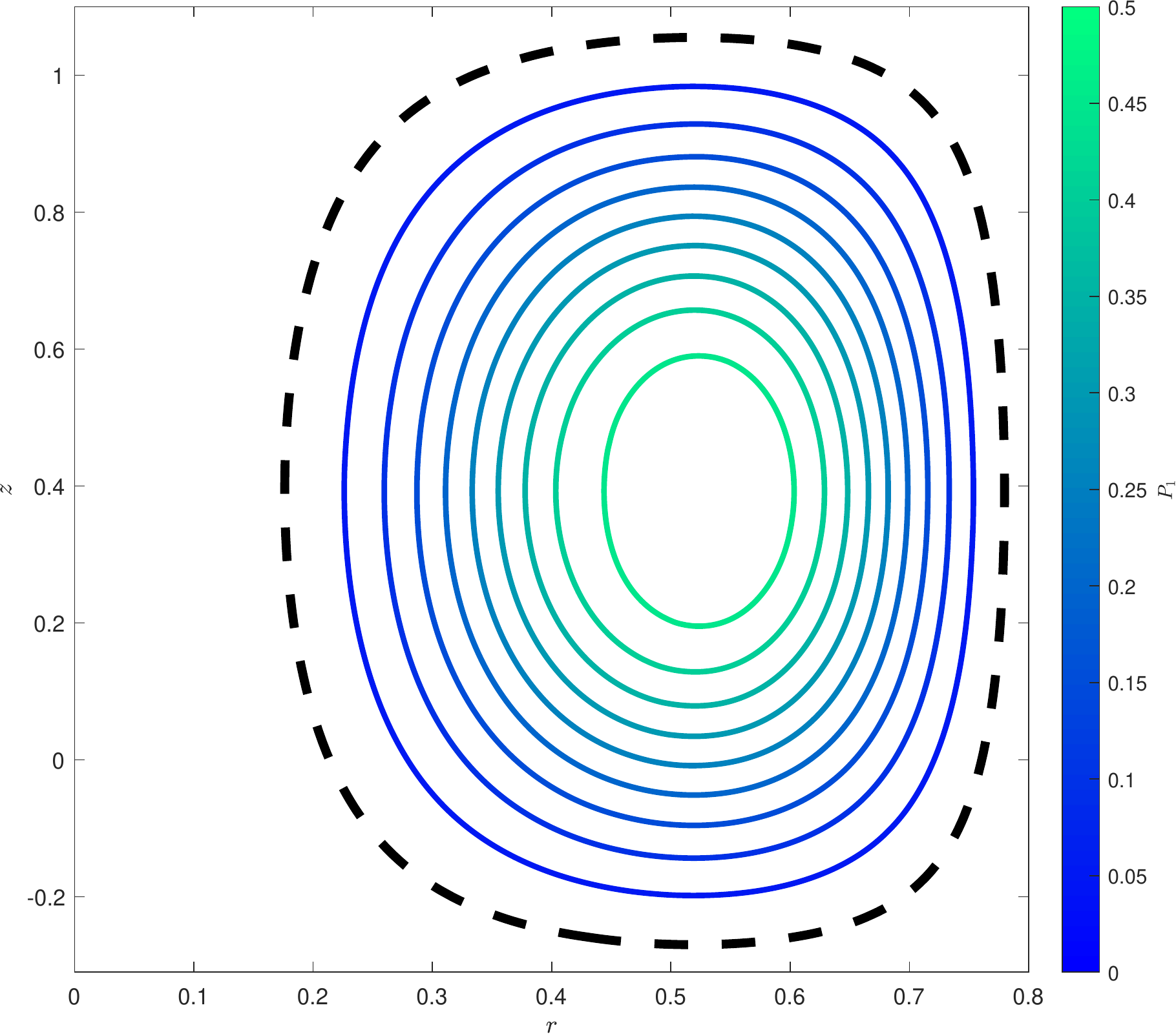}
\end{center}
\caption{\label{fig:Axial_family2_pressure_contour_transform}A cross-section of magnetic surfaces $P_1 = \const$ for a sample axially symmetric plasma equilibrium solution belonging to a Family 2, \eqref{Sep_solution_2}, with $C_1 = 1$, $C_2 = 0$, $C_3 = 1$, $C_4 = 1$, $k = 2$, $\alpha = 5$, and $q = \sqrt{3}$ after the transformation with \eqref{Bog_trans_st} with $F_1 = 0.7P^4$ and $F_2 = P^2$. The truncated boundary is shown boldface which coincides with the current sheet marking the boundary of the plasma.}
\end{figure}

\begin{figure}[htbp]	
\begin{center}
\includegraphics[width = 1\textwidth]{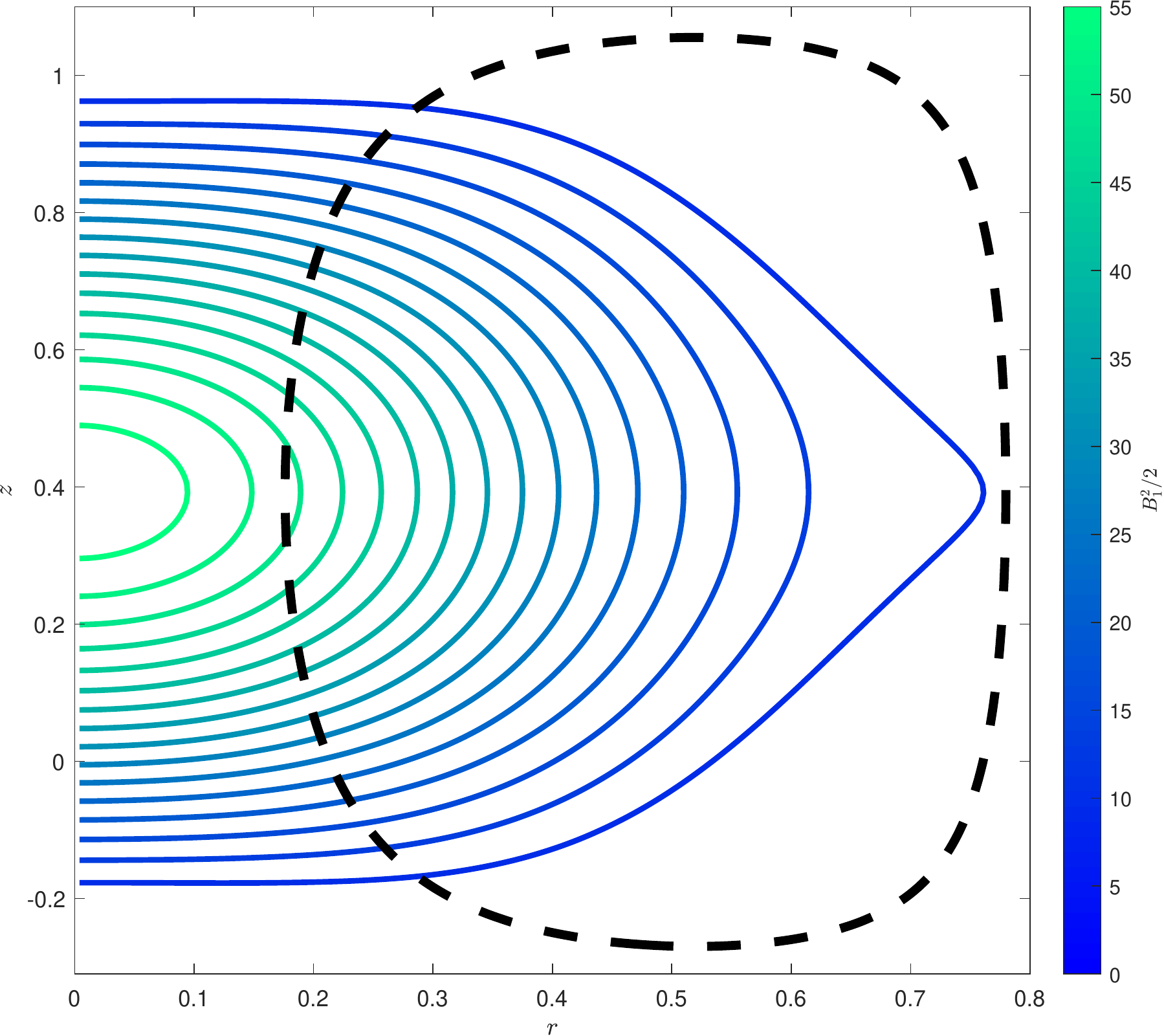}
\end{center}
\caption{\label{fig:Axial_family2_Mag_energy_density_transform}Lines of constant Magnetic energy density $B_1^2/2$. This comes from an axially symmetric plasma equilibrium solution belonging to a Family 2, \eqref{Sep_solution_2}, with $C_1 = 1$, $C_2 = 0$, $C_3 = 1$, $C_4 = 1$, $k = 2$, $\alpha = 5$, and $q = \sqrt{3}$ after the transformation with \eqref{Bog_trans_st} with $F_1 = 0.7P^4$ and $F_2 = P|P|$.}
\end{figure}

\begin{figure}[htbp]	
\begin{center}
\includegraphics[width = 1\textwidth]{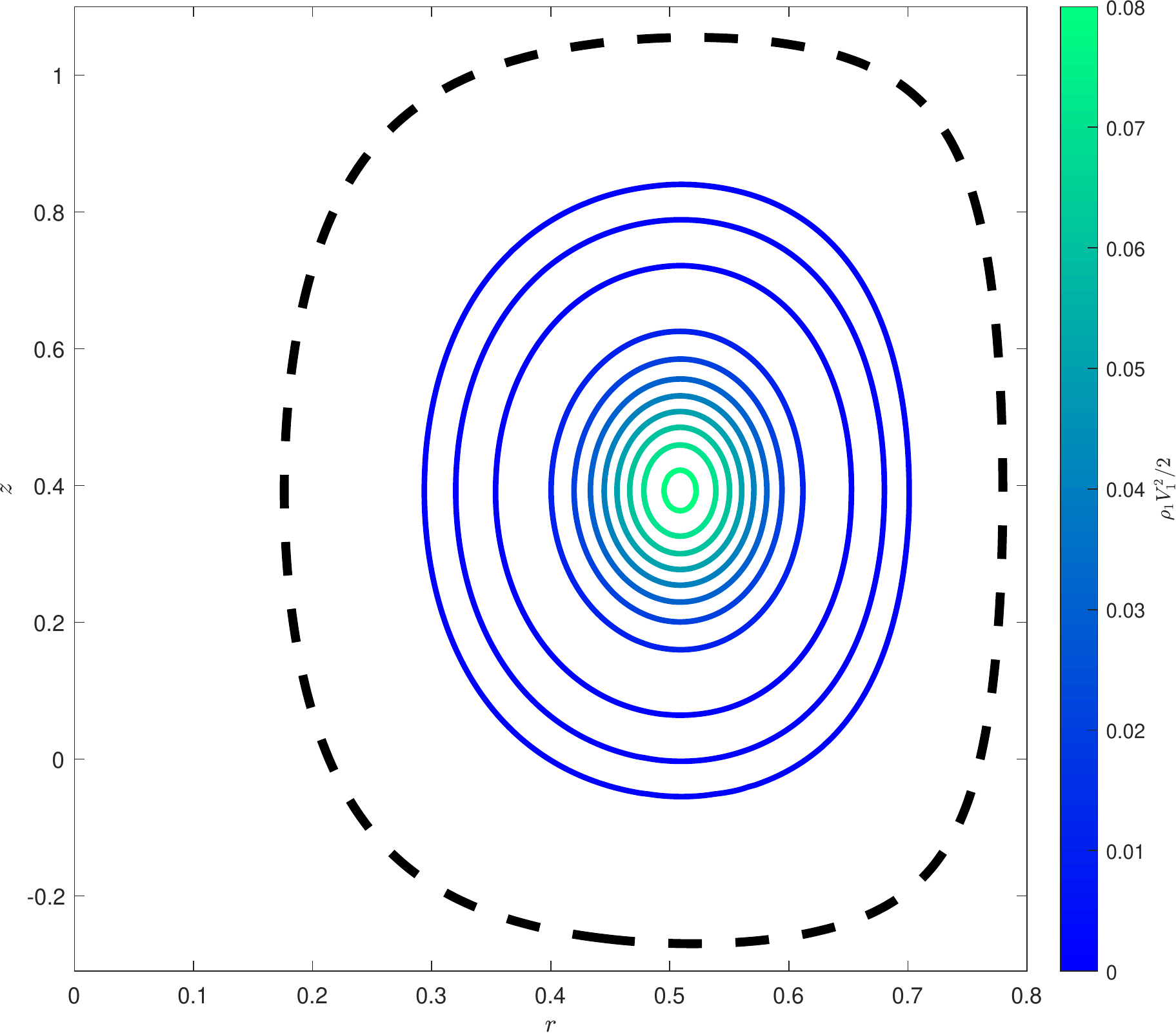}
\end{center}
\caption{\label{fig:Axial_family2_kinetic_energy_density_transform}Lines of constant kinetic energy density . This comes from an axially symmetric plasma equilibrium solution belonging to a Family 2, \eqref{Sep_solution_2}, with $C_1 = 1$, $C_2 = 0$, $C_3 = 1$, $C_4 = 1$, $k = 2$, $\alpha = 5$, and $q = \sqrt{3}$ after the transformation with \eqref{Bog_trans_st} with $F_1 = 0.7P^4$ and $F_2 = P|P|$. The colorbar shows the values of the dimensionless kinetic energy density $\rho_1V^2_1/2$}
\end{figure}

Starting with the above example for the second family of axially symmetric solutions shown in \Figref{Axial_Coulomb_contour_pressure}, having the magnetic field components and pressure given as
\begin{subequations}
\begin{equation}
\vec{B}_{st} = \frac{\psi_r}{r}\vec{e}_r + \alpha \frac{\psi}{r} \vec{e}_\varphi - \frac{\psi_z}{r}\vec{e}_z,
\end{equation}
\begin{equation}
P = P_0 + \frac{q^2}{2}\psi^2,
\end{equation}
\end{subequations}
where $\psi$ is given by (\ref{eq:Sep_solution_2}) and using the two arbitrary functions of $\psi$ to be
\begin{equation}
F_1(\psi) = 0.7P^4,\quad
F_2(\psi) = P^2,
\end{equation}
and $C = 1$, this transforms the solution into
\begin{equation*}
\vec{B_1} = \sqrt{1 + P^4}\vec{B}_{st}, \quad \vec{V_1} = \frac{\vec{B}_{st}}{0.7P^2},
\end{equation*}
\begin{equation}\label{eq:trans1}
 \quad P_1 = P - \frac{P^2\vec{B}_{st}^2}{2}, \quad \rho_1 = P^4.
\end{equation}
The pressure profile and magnetic energy density can be seen in \Figref{Axial_family2_pressure_contour_transform} and \Figref{Axial_family2_Mag_energy_density_transform} respectively. With the non-zero velocity, a graph of the non-zero kinetic energy density, $\rho_1 V_1^2/2$, can now be seen and is shown in \Figref{Axial_family2_kinetic_energy_density_transform}.

\subsection{Transformations of Helical family 2 example}

\begin{figure}[htbp]	
\begin{center}
\includegraphics[width = 1\textwidth]{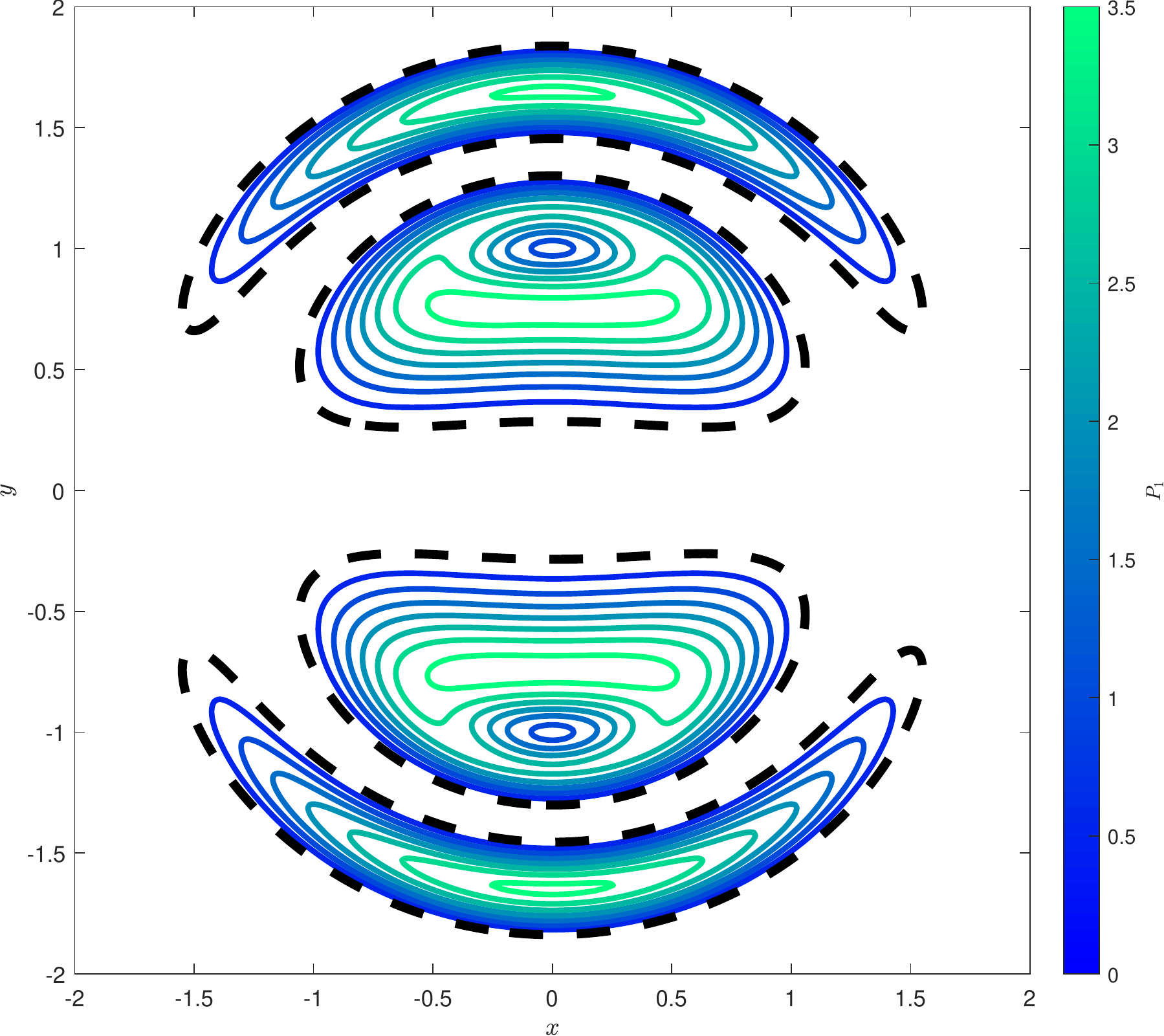}
\end{center}
\caption{\label{fig:Helical_family2_pressure_contour_transform}A cross-section of the new magnetic surfaces $ P1 = \const$ for a sample helically symmetric plasma equilibrium solution belonging to Family 2 using \eqref{Sep_solution_4} with $C_1 = 1$, $C_2 = 0$, $\alpha = 3$, $\kappa = 4$, $\gamma = 1$ and $\omega = 1$ after transforming using \eqref{Bog_trans_st} with $F_1 = 0.035P^2$ and $F_2 = \cos(\psi^2)e^{\frac{\psi^4}{2}}$. The colorbar shows the values of the dimensionless pressure $P_1$. The truncated boundary is shown black and dashed which coincides with the current sheet marking the boundary of the plasma.}
\end{figure}

\begin{figure}[htbp]	
\begin{center}
\includegraphics[width = 1\textwidth]{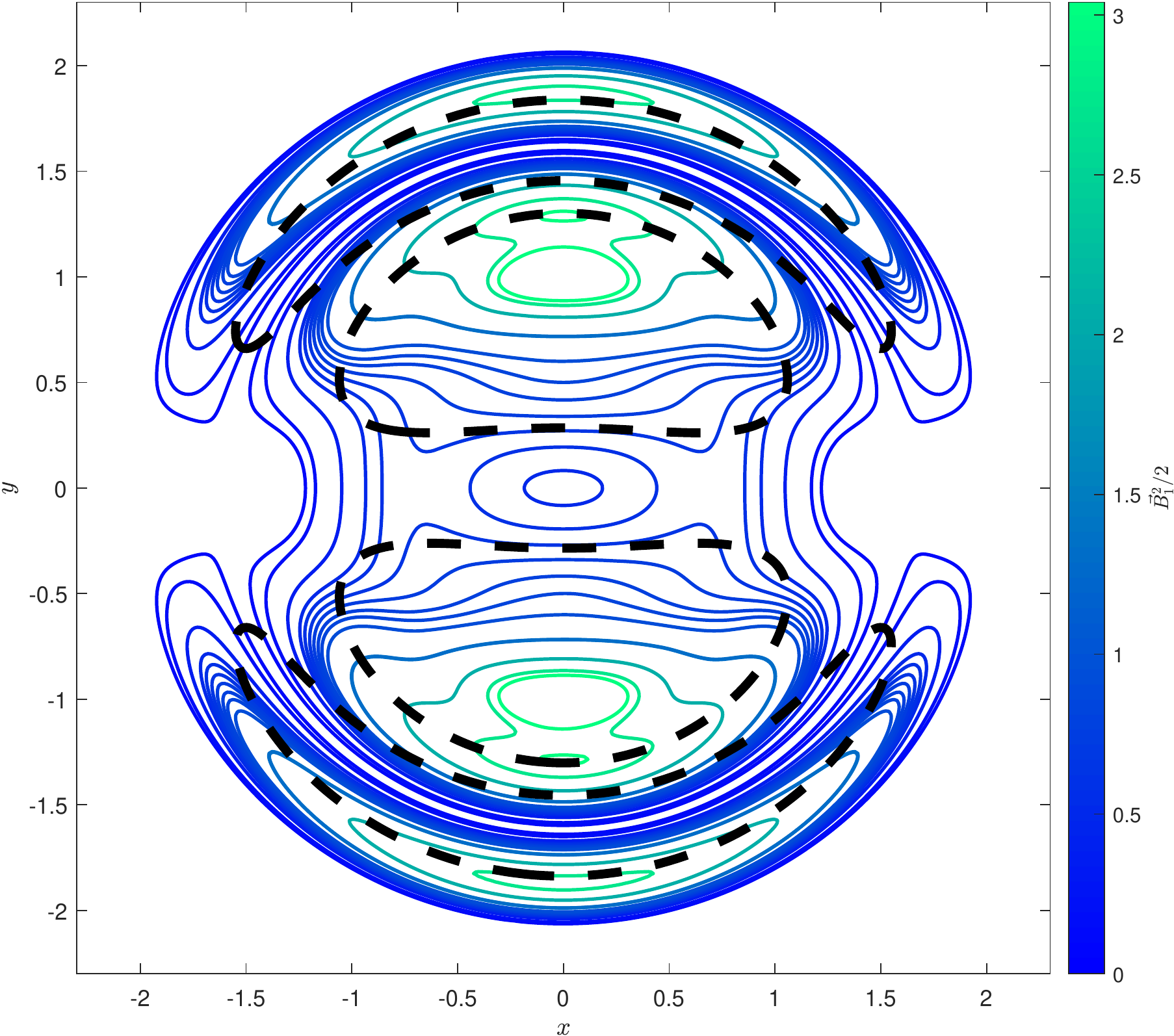}
\end{center}
\caption{\label{fig:Helical_family2_Mag_energy_density_transform}Lines of constant magnetic energy density. This comes from a Helically symmetric plasma equilibrium solution belonging to Family 2 using \eqref{Sep_solution_4} with $C_1 = 1$, $C_2 = 0$, $\alpha = 3$, $\kappa = 4$, $\gamma = 1$ and $\omega = 1$ after transforming using \eqref{Bog_trans_st} with $F_1 = 0.035P^2$ and $F_2 = \cos(\psi^2)e^{\frac{\psi^4}{2}}$. The colorbar shows the values of the dimensionless magnetic energy density, $B_1^2/2$. The thick black dotted line marks the boundary of the plasma with the area of interest being inside of this boundary.}
\end{figure}

\begin{figure}[htbp]	
\begin{center}
\includegraphics[width = 1\textwidth]{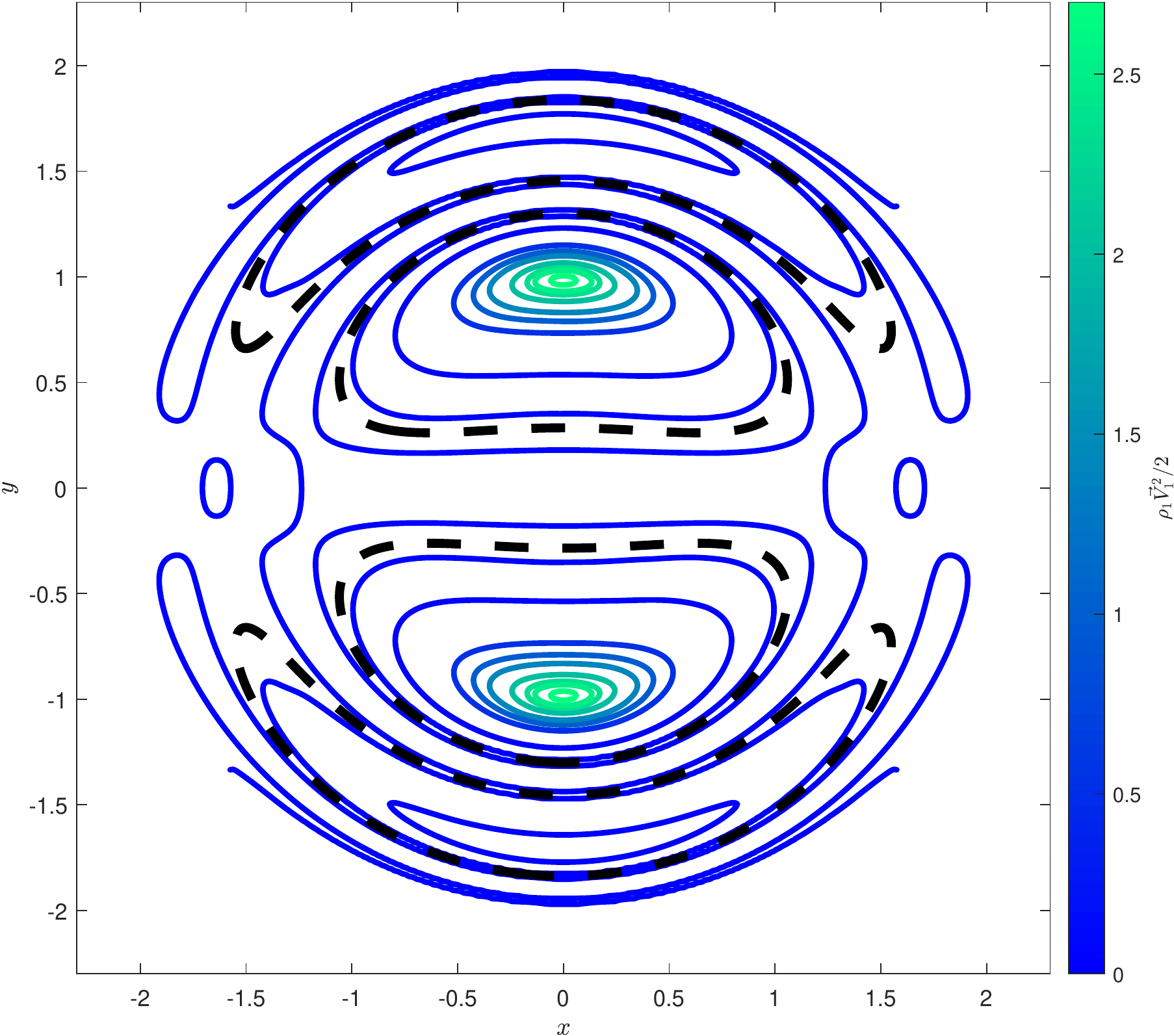}
\end{center}
\caption{\label{fig:Helical_family2_kinetic_energy_density_transform}Lines of constant kinetic energy density. This comes from a Helically symmetric plasma equilibrium solution belonging to Family 2 using \eqref{Sep_solution_4} with $C_1 = 1$, $C_2 = 0$, $\alpha = 3$, $\kappa = 4$, $\gamma = 1$ and $\omega = 1$ after transforming using \eqref{Bog_trans_st} with $F_1 = 0.035P^2$ and $F_2 = \cos(\psi^2)e^{\frac{\psi^4}{2}}$. The colorbar shows the values of the dimensionless kinetic energy density, $\rho_1 V_1^2/2$. The thick black dotted line marks the boundary of the plasma with the area of interest being inside of this boundary.}
\end{figure}

Taking the second family of helically symmetric solutions shown in \Figref{Helical_family2_pressure_contour}, having the magnetic field components and pressure given as
\begin{subequations}
\begin{equation}
\vec{B}_{st} = \frac{\psi_\xi}{r}\vec{e}_r + \frac{\alpha r \psi + r\psi_r}{r^2 + \gamma^2} \vec{e}_\varphi + \frac{\gamma \alpha \psi - r\psi_r}{r^2 + \gamma^2}\vec{e}_z,
\end{equation}
\begin{equation}
P = P_0 + \frac{\kappa^2}{2}\psi^2,
\end{equation}
\end{subequations}
where $\psi$ is given by (\ref{eq:Sep_solution_4}) and using the two arbitrary functions of $\psi$ to be
\begin{equation}
F_1(\psi) = 0.035 P^2,\quad
F_2(\psi) = \cos(\psi^2)e^{\frac{\psi^4}{2}}.
\end{equation}
and $C = 1$, this transforms the solution into
\begin{equation*}
\vec{B_1} = \sqrt{1 + \cos^2(\psi^2)e^{\psi^4}}\vec{B}_{st}, \quad \vec{V_1} = \frac{\cos(\psi^2)e^{\frac{\psi^4}{2}} }{0.035P^2}\vec{B}_{st},
\end{equation*}
\begin{equation}\label{eq:trans2}
P_1 = P - \frac{\cos^2(\psi^2)e^{\psi^4}}{2}\vec{B}_{st}^2, \quad \rho_1 = \cos^2(\psi^2)e^{\psi^4}.
\end{equation}
The new pressure contour and magnetic energy density can be seen in \Figref{Helical_family2_pressure_contour_transform} and \Figref{Helical_family2_Mag_energy_density_transform} respectively. The new non-zero kinetic energy density is shown in \Figref{Helical_family2_kinetic_energy_density_transform}.

\section{Discussion}

In this paper, exact physical solutions to the system of MHD equations \eqref{MHD} were sought after. Two separate symmetry reductions were performed on the static equilibrium MHD equations (\ref{eq:MHDst}). These two reductions correspond to two different symmetries of plasma found in nature. These include axial invariance and helical invariance.\medskip

For the axial invariance reduction, one can introduce a potential flux function $\psi(r,z)$ which reduces the system of PDEs given by (\ref{eq:MHDst}) into the Grad-Shafranov (\ref{eq:GS}). The arbitrary functions $P(\psi)$ and $I(\psi)$ can be chosen with (\ref{eq:P_I_axial_linear}) so that the G-S equation becomes linear, (\ref{eq:GS_lin}). From here, separation of variables can be used to find solutions. Two separate families of solutions arise depending on the type of pressure configuration chosen. For pressure in which the plasma is surrounded by atmosphere, the radial component has solutions in terms of Whittaker functions. The exact solution of this type is given by \eqref{Sep_solution1}. Due to the behaviour of Whittaker functions, if the first parameter $\delta$ of these functions is a non-integer value only truncated solutions can be considered physical, where the chosen boundary is described with the surface current sheet given by (\ref{eq:Boundary_Condition}). However, when $\delta$ is an integer value, any linear combination of solutions can be considered physical without truncation. It is proven in the appendix that these are the only physical solutions to (\ref{eq:GS_lin}) which do not require truncation at some boundary. An example of a truncated solution from this family has a pressure profile seen in \Figref{axial_whittaker_pressure_plot} and the corresponding magnetic energy density seen in \Figref{Axial_Whittaker_mag_energy_plot}. A solution which doesn't require truncation was also plotted with magnetic surfaces and magnetic energy density given by \Figref{whittaker_bog_contour_pressure} and \Figref{whittaker_bog_contour_mag_energy} respectively.\medskip

The second family of solutions for the axial case has the radial component written in terms of Coulomb wave functions. The exact solution of this type is given by \eqref{Sep_solution_2}. Magnetic surfaces with the positive pressure profile (pressure higher in the center of the plasma) can be seen by \Figref{Axial_Coulomb_contour_pressure} along with its corresponding magnetic energy density seen in \Figref{Axial_family2_mag_energy_density}. Due to the periodic behaviour of the Coulomb wave functions, for physical solutions, they too must be truncated at some chosen plasma boundary. \medskip

For the helical case, after introducing a potential flux function $\psi(r,\xi)$ the system of PDEs (\ref{eq:MHDst}) reduces to the JFKO equation (\ref{eq:JFKO}). After The arbitrary functions $P(\psi)$ and $I(\psi)$ are chosen with (\ref{eq:P_I_helical_linear}), the JFKO equation becomes linear, (\ref{eq:JFKO_lin}). Similarly to the axial case, using separation of variables, two different families of solutions arise depending on the type of pressure configuration. For plasmas confined in atmosphere, the solution are given by \eqref{Sep_solution_3}. An example not requiring truncation is shown with pressure profile and magnetic energy density seen in \Figref{Helical_bog_contour_pressure} and \Figref{Helical_bog_mag_energy_density} respectively. There is a special case of this solution, where the confluent Heun function produces polynomials. These types of solutions are given by \eqref{helical_jets} and have pressure profile seen in \Figref{Helical_bog_contour_pressure} with the corresponding magnetic energy density \Figref{Helical_bog_mag_energy_density}.\medskip

The second helical family of solutions corresponding to a plasma confined in vacuum has solution given by \eqref{Sep_solution_4}. The radial part of this solution is written in terms of confluent Heun functions with imaginary parameters and the behaviour is similar to that of the Coulomb wave functions. Therefore this family of solutions must be truncated to be a physical solution. The truncated positive pressure profile can be seen in \Figref{Helical_family2_pressure_contour} with the corresponding magnetic energy density \Figref{Helical_family2_magnetic_energy}.\medskip

In the last section of this paper, the positive pressure configurations for both the axial and helical solutions were transformed into solutions with $\vec{V} \ne 0$. The new magnetic surfaces, magnetic energy density and kinetic energy density are shown in \Figref{Axial_family2_pressure_contour_transform},  \Figref{Axial_family2_Mag_energy_density_transform}, and \Figref{Axial_family2_kinetic_energy_density_transform} respectively for the axial case and similarly \Figref{Helical_family2_pressure_contour_transform}, \Figref{Helical_family2_Mag_energy_density_transform} and \Figref{Helical_family2_kinetic_energy_density_transform} respectively for the helical case. \medskip

One open problem with the first family of helical solutions which is put forward is finding the necessary conditions for the parameters in \eqref{Sep_solution_3} to produce solutions which do not need to be truncated to satisfy the physical requirements.

\begin{appendix}

\section{Proof of proposition 1}\label{appendix:a}

\emph{If $\delta \notin \mathbb{N}$, \eqref{GS_lin} must be truncated to be a physical solution.}
\medskip

Assuming $\delta \notin \mathbb{N}$. The general $R(r)$ solution given by (\ref{eq:Whittaker_Solution}) with $C_1 = 1$ and $C_2 = 0$, for a large argument, $r \to \infty$ can be approximated by \cite{NIST:DLMF} as

\begin{equation}
W_M(\delta,1/2,qr^2) \approx \frac{\Gamma(2)}{\Gamma({1 - \delta})}e^{\frac{qr^2}{2}}(qr^2)^{-\delta}.
\end{equation}

Next, the limit of this at $\infty$ is considered, which is related to the well known limit $\displaystyle{\lim_{x \to \infty} \frac{e^x}{x^k}}$ which is known to diverge to $\infty$, giving

\begin{equation*}
\lim_{r \to \infty} W_M(\delta,1/2,qr^2) \to \infty.
\end{equation*}

Therefore, the axial magnetic field component which behaves like $W_M(\delta,1/2,qr^2)/r$ in the $r$ variable will also diverge at infinity,

\begin{equation*}
\lim_{r \to \infty}{B^{\phi}}  \to \infty.
\end{equation*}

For $C_1 = 0, C_2 = 1$, and small $r$, the approximation of $W_W(\delta,1/2,qr^2)$ given by \cite{NIST:DLMF} can written as

\begin{equation*}
W_W\left(\kappa,\frac{1}{2},qr^2\right) \approx \frac{1}{\Gamma(1 - \delta)},
\end{equation*}

which is simply a non-zero constant term which are well defined as the argument of the Gamma function will never be negative half integers. Therefore, the value of $B^\phi(r,z)$ which behaves like $W_W(\delta,1/2,qr^2)/r$ in the $r$ variable will diverge at the origin.

\begin{equation*}
\lim_{r \to 0}{B^{\phi}}  \to \infty.
\end{equation*}

Clearly any linear combination of $W_M(\delta,1/2,qr^2)$ and $W_W(\delta,1/2,qr^2)$ will not have finite $B^{\phi}$ for $0 \leq r < \infty$, so this solution cannot be physical without truncation.\\

\emph{if $\delta \in \mathbb{N}$, \eqref{GS_lin} doesn't need to be truncated to be a physical solution.}
\medskip

Assuming $\delta \in \mathbb{N}$, then $W_M(\delta,1/2,qr^2)$ and $W_W(\delta,1/2,qr^2)$ in  (\ref{eq:Whittaker_Solution}) become linearly dependent and behave like
\begin{equation}
W_M(\delta,1/2,qr^2) \propto W_W(\delta,1/2,qr^2) \propto r^2e^{(-\frac{qr^2}{2})}L_{\delta - 1}(qr^2),
\end{equation}

where $L_{\delta - 1}$ are the $\delta - 1$ order Laguerre polynomials. Using (\ref{eq:Axial_Magnetic_Field}), clearly the $B^{\phi}$ and $B^{r}$ which behave like $r e^{(-\frac{qr^2}{2})}L_{\delta - 1}(qr^2)$ in the r variable, are smooth, finite in the interval $0\leq r <\infty$ and go to zero for $r \to \infty$ as this is simply the product of a polynomial and Gaussian, using the well-known result that $\displaystyle{\lim_{r \to \infty} {P(r)}{e^{-r^2}} = 0}$, where $P(r)$ is any polynomial in $r$. Therefore finite magnetic energy will be the case with these components. For the last component, $B^{z}$, from (\ref{eq:Axial_Magnetic_Field}), the first derivative in r must be checked,

\begin{equation}\label{eq:proof2}
\frac{d}{dr} \left(r^2e^{(-\frac{qr^2}{2})}L_{\delta - 1}(qr^2)\right) = -r e^{(-\frac{qr^2}{2})}\left(2\delta L_{\delta-1}(qr^2) + (qr^2 - 2\delta - 2))L_\delta(qr^2)\right)
\end{equation}

Clearly then $B^z$ which behaves like $-e^{(-\frac{qr^2}{2})}\left(2\delta L_{\delta-1}(qr^2) + (qr^2 - 2\delta - 2))L_\delta(qr^2)\right)$ in the r variable is also smooth, finite in the interval $0\leq r <\infty$ and goes to zero for $r \to \infty$ again since this is the product of a Gaussian and polynomial. Therefore the quantity
\begin{equation}
\int_{\mathcal{U}} |\vec{B}(\vec{x})|^2\, d^3 x
\end{equation}
will be finite. Also, the pressure $P(\psi) = P_0 -  1/2\psi^2 \to P_0$ for $r \to \infty$ since $\psi = R(r)Z(z) \to 0$ for $r \to \infty$, so the second physical requirement is also satisfied.

\end{appendix}

\subsubsection*{Acknowledgements}

The authors are grateful to NSERC of Canada for the financial support.
%%

%\printbibliography
{\footnotesize
\bibliography{MHDKellerv04}
\bibliographystyle{ieeetr}
}

\end{document}